\documentclass[preprint,12pt]{revtex4-1}

\usepackage{graphicx}
\usepackage{dcolumn}
\usepackage{bm}
\usepackage{lipsum}
\usepackage{amssymb}
\usepackage{float}
\usepackage[caption = false]{subfig}
\usepackage{amsmath}
\usepackage{amsfonts}
\usepackage{amssymb}
\usepackage{amsmath}
\usepackage[section]{placeins}
\usepackage{xcolor}

\usepackage{enumitem}

\newcommand{\R}{I\hspace{-1.2mm}R}
\newcommand{\N}{I\hspace{-1.2mm}N}

\begin{document}

\title{Distribution Probability of Force for a  Physical System of $N$ Random Particles\\{\em{(to appear in Journal of Mathematical Physics)}}}

\author{A.~D.~Figueiredo}
\affiliation{Instituto de F\'{i}sica and International Center for Condensed Matter Physics,
Universidade de Bras\'\i lia, CP 04455, 70919-970 - Bras\'\i lia, Brazil}

\author{T.~M.~da~Rocha Filho}
\affiliation{Instituto de F\'{i}sica and International Center for Condensed Matter Physics,
Universidade de Bras\'\i lia, CP 04455, 70919-970 - Bras\'\i lia, Brazil}

\author{M.~A.~Amato}
\email{maamato@unb.br}
\affiliation{Instituto de F\'{i}sica and International Center for Condensed Matter Physics,
Universidade de Bras\'\i lia, CP 04455, 70919-970 - Bras\'\i lia, Brazil}



\begin{abstract}
The present paper  attempts to address a discussion on mathematical grounds of a model to associate the  generalized version of the CLT and the $N$-body problem related to the calculation of the force on a single star or particle due to the $N-1$ stars or particles whenever they are randomly distributed in the space and $N\rightarrow\infty$. We calculate the resultant force on a test particle immersed in a $N$-particle system under a $1/r^\delta$ force ($\delta>0$) and discuss the limit force under different approaches referred to as the Vlasov limit and Fluctuation Limit. Also one shows the behaviour of the limit force in different domains for the L\'evy exponent ($\alpha$). 

{\em {This paper has been accepted to appear in JOURNAL OF MATHEMATICAL PHYSICS}}

\end{abstract}

\maketitle

\section{Introduction}

The work of L\'evy on non Gaussian stable laws \cite{levy4} allows the interpretation of the limit sum of $N$ real random variables as a one-dimensional resultant force from $N$ particles on a test particle. For that the particles are uniformly distributed in space with a pairwise interaction and the force with a power law dependence. 
Subsequently, this result has been generalized in~\cite{levy3} and may be interpreted as the resultant force of an arbitrary distribution of particles in an infinity line. The extension to built the limit theory for the sum of random vectors was first made by Levy in~\cite{levy1}-\cite{levy2} and later using the famous tringular renormalization scheme of Kolmogorov \cite{gnedenko} by Rvaceva in~\cite{rvaceva} and this work has been the basis for further developments in the study of Limit Theorems for the sum of random vectors.

L\'evy~\cite{levy4}-\cite{levy2} and Levy and Khintchine~\cite{levykhintchine} have demonstrated rigorously that the stable laws are infinitely divisible and therefore may obtained as a limit of a sum of identically distributed random vectors. Such result relies on the direct extension of the theory of infinitely divisible laws for real random variables. 

The mathematical form to represent the characteristic function associated with a given stable law most used and also considered standard is the representation due to Samorodtsky and Taqqu~\cite{taqqu,hall,mcculloch}.

In order to determine for which stable law the sum process of random vectors converges, the most important property of a probability density is its asymptotic behaviour at infinity. According to L\'evy (see~\cite{levy1}) it is required that the probability density should have an  asymptotic behaviour as a power law, or at least, bounded by a power law. More recently, using the notion of functions with regular variation \cite{karamata1}--\cite{karamata2}  and according to Feller \cite{feller}, it was introduced a more general characterization of asymptotic properties of a probability density implying convergence to stable laws \cite{balkema}--\cite{meerschaert3}. A more comprehensive mathematical treatment on the problem of limit distributions for sums of random vectors can be found in~\cite{meerschaert4}. 

In physics due to the importance of the understanding of the statistical properties of the gravitational force, the pioneer studies of Chandrasekhar~\cite{chandra}, Antonov~\cite{antonov} and LyndenBell and Wood~\cite{lyndenbell} on the thermodynamics and dynamics of galaxies and stars led to the development of a set of very similar methods to those developed for the  Central Limit Theorem (CLT) although with a limited scope and aim. Hereafter, $d$ and $\delta$ stand for dimensionality of the system and power law of the force, respectively.

Chandrasekhar~\cite{chandra} develops along the same reasoning of the work of Holtsmark~\cite{holt} in the context of plasma physics for gravitational force ($d=3$, $\delta=2$) and obtains the limit distribution for the resultant force on a test particle located in the centre of a sphere of radius $L$ for a system of $N$ uniformily distributed particles inside the sphere ($d$ stands for the dimension of the system). The result obtained in the thermodynamic limit with $N,L\rightarrow\infty$ and $3N/4\pi L^3=const$ is a symmetrical stable L\'evy distribution with $\alpha=3/2$ which is the same exponent $\left(\alpha=\frac{3}{2}\right)$ of the Holtsmark distribution. Ahmad and Cohen~\cite{ahm} studied the same system with $N\rightarrow\infty$ but a finite $L$ and could not obtain a limit distribution for the resultant force. Instead, they obtained an assymptotic expression for the force as a function of $N$. Kandrup~\cite{kandrup,kandrup1} extends these results for an isotropic and non uniform distribution and notice that whatever the size of sphere ($L$) it is unimportant for the determination of the distribution of the resultant force in its assymptotic form.

In a set of papers Chavanis and Sire~\cite{chav0,chav1} and Chavanis~\cite{chav2, chavanis} studied the problem in different dimensions ($d=2$, $\delta=1$)) and ($d=1,\delta=0$)) along the same lines of the theories developed for $d=3$ systems. For$d=2$ and a uniform distributuion, in the thermodynamic limit $N, L\rightarrow\infty$ and $N/\pi L^2=constant$ the results show that one cannot obtain a well defined limit and assymptocally the distribution has a gaussian behaviour but its second moment diverges with $N$. For $d=1$, representing an effective gravitional force between infinity planes,the particle are distributed simmetrically in an interval of size $2L$, with the test particle at the midpoint of the interval, the results show that a proper thermodynamic limit ($N/L=const$) is absent. In this case he obtains an assymptotic expression for the distribution of the resultant force but with first and second moments proportional to the system size ($N$). In~\cite{chavanis} Chavanis extends the results for a dimension $d$ and $\delta>0$, an isotropic distribution ($\rho(r)=C r^\nu$ ($C>0$) for a system of $N$ particles in a sphere of size $L$ with a test particle at its centre. Imposing the restrictions $-d<\nu\leq 0$ and $0<\alpha=(d+\nu)/\delta<2$ he obtains the limit distribution of the resultant force for $N,L\rightarrow\infty$ with the constraint $NL^{d+\nu}=constant$. In this conditions the distribution is a L\'evy symmetrical and stable distribution with exponent $\alpha=(d+\nu)/\delta$. 

Amid this amount of work on the subject one notices that a sistematic study of the conditions for the existence of a well defined limit for the resultant force is still necessary. To carry out this study one makes use of the mathematical tools developed by L\'evy in his approach to obtain the Central Limit Theorem. The present paper  attempts to address to this fundamental problem in which gravitational forces play a central role and for that one discuss on a more rigorous mathematical ground a model to associate the  generalized version of the CLT and the $N$-body problem related to the calculation of the force on a single star or particle due to the $N-1$ stars or particles whenever they are randomly distributed in the space and $N\rightarrow\infty$. On this grounds we have obtained in more general terms the conditions for the existence of this limit and not needing to impose condition of either uniformity or isotropy for the distribution of $N$ particles that act on the test particle. This study encompass the previous results presented in the literature and are shown in Tables~\ref{tab5}~and~\ref{tab4}. A thorough revision of these problems may be found in~\cite{padma} and~\cite{chavanis0}. 

The paper has the following structure: in Section~\ref{clt} we discuss the CLT for the sum of random force and provide the mathematical tools in the context of probability theory, in Section~\ref{LK} one obtains the L\'evy-Khintchine Stable Distributions of the Resultant Force, in Section~\ref{prob} we calculate the resultant force on a test particle immersed in a $N$-particle system under a $1/r^\delta$ force ($\delta>0$) and discuss the limit force under different approaches referred as to Vlasov limit and Fluctuation Limit. Also one shows the behaviour of the limit force in different domains for the L\'evy exponent ($\alpha$). In Section~\ref{conclusion} we present our concluding remarks about the process of renormalization. In the appendices some stages of the mathematical results are included in order to make the paper more self contained and some applications are also presented.

\section{Central Limit Theorem for the Sum of Random Forces}
\label{clt}

The main issue related to the general version of the CLT is to characterize the probability density of a given random vector, in such a way that the sum of $N$ of these random vectors converge to a stable law when $N\rightarrow\infty$. This sum can  be done by a proper normalization and centralization of the random vector corresponding to the sum of these $N$ identical vectors. This process will be referred to  as Renormalization.

In this section we aim to apply the CLT to calculate the net force on a given particle, called {\it Test Particle}, due to its interaction with others $N$ identical particles and the force of each one of these  particles on the test particle is supposed to be described by a random vector with a power law density distribution. For this we make use of the  CLT for a sum of identical random forces that is directly based on the formalism described in \cite{levy4} and \cite{levy3} for sum of the random real variables \cite{levy5}.

\subsection{Basic Definitions and Notations}

A $d$-dimensional real random vector $\vec{U}$ is an ordered collection of $d$ real random variables denoted by a capital letter with an arrow and its components by the same letter with a index subscript and represented by $\vec{U}=\left(U_1,\ldots,U_d\right)$. We admit in the following that all $U_i$'s are continuous random variables. Although the mathematical results presented in this section  remain valid for any dimension $d$, for the random force in the physical space one must should restrict to $d=1,2,3$ dimensions. 

The probability density $\rho_{\vec{U}}(\vec{u})$ of a simultaneous realization $\vec{u}=(u_1,\ldots,u_d) $ of each component $U_i$ of $\vec{U}$ satisfy the usual criteria, i.e., $\rho_{\vec{U}}(\vec{u})\geq 0$ and  $\int_{\R^d}\rho_{\vec{U}}(\vec{u})d^d\vec{u}=1$. The mean of a function $f(\vec{U})$ and the characteristic function of $\rho_{\vec{U}}(\vec{u})$ are defined by:

\begin{equation}
\left<f(\vec{U})\right>=
\int_{\R^d}f(\vec{u})\rho_{\vec{U}}(\vec{u})d^d\vec{u},
\end{equation}

\begin{equation}
\psi_{\vec{U}}(\vec{z})=
\left<\exp\left(i\vec{z}\cdot\vec{u}\right)\right>=\int_{\R^d}\exp\left(i\vec{z}\cdot\vec{u}\right)\rho_{\vec{U}}(\vec{u})d^d\vec{u},
\end{equation}

\noindent respectively, with $\vec{z}\in\R^d$ and $\vec{z}\cdot\vec{u}\equiv z_1u_1+\cdots+z_nu_d$.
The important point is that there is a one to one correspondence between density distributions and characteristic functions.
If the second order statistical moments of $\vec{U}$ exists, then we can define the covariance matrix of a random vector $\vec{U}$
as the $d\times d$ matrix $\bar{M}$ with components:

\begin{equation}
\bar{M}_{ij}=\left<U_iU_j\right>-\left<U_i\right>\left<U_j\right>.
\label{equ1}
\end{equation}

On these grounds the following properties hold:

\begin{enumerate}
\item {\it The renormalization property}: For the afine transformation $\vec{W}=a\vec{U}+\vec{b}$ of the random vector $\vec{U}$,
with $a$ being a real number and $\vec{b}\in\R^d$, we have:
$$\psi_{\vec{W}}\left(\vec{z}\right)=\exp\left(i\vec{b}\cdot\vec{z}\right)\psi_{\vec{U}}\left(a\vec{z}\right),$$
\item {\it The convolution property}: For $\vec{U}$ and $\vec{W}$ being two independent random vectors, the characteristic function of $\vec{U}+\vec{W}$ is given by:
$$\psi_{\vec{U}+\vec{W}}\left(\vec{z}\right)=
\psi_{\vec{U}}\left(\vec{z}\right)\psi_{\vec{W}}\left(\vec{z}\right).$$
\item {\it L\'evy's Continuity Theorem}  For any sequence of random vectors $\vec{F}_k$, $k\in\N$, such that the sequence of their corresponding characteristic
functions converges continuously to some function $\phi(\vec{z})$, that is:
$$\lim_{k\rightarrow\infty}\psi_{\vec{F}_k}(\vec{z})=\phi(\vec{z}),$$
then the function $\phi(\vec{z})$ is a the characteristic function of a random vector.
\end{enumerate}

\subsection{Random Force Distribution with Power Law Tails}

Consider a $d$-dimensional random force $\vec{U}$, with density probability $\rho_{\vec{U}}(\vec{u})$ given by:

\begin{equation}
\label{equ2}
\rho _{\vec{U}} \left(\vec{u}\right)=
\left\{
\begin{array}{cc} 
f\left(\vec{u}\right) & {\rm if}\hspace{5mm}\left|\vec{u}\right|\le u_c,\\
 & \\ \displaystyle \frac{C\left(\hat{u}\right)}{\left|\vec{u}\right|^{d+\alpha}}  & {\rm if}\hspace{5mm}\left|\vec{u}\right|>u_c
\end{array}
\right.
\end{equation}

\noindent with $u_c>0$, $\alpha>0$, $C(\hat{u})$ any non-negative function defined on the hypersphere of unit radius, $S_d$ ,
and $f(\vec{u})$ a positive function defined for $|\vec{u}|\leq u_c$. The functions $f(\vec{u})$
and $C(\hat{u})$ must be such that $\rho_{\vec{U}}(\vec{u})$ is a normalized distribution.
In order to show that the characteristic function of a power law distribution of random vectors can be put in the canonical form  one has to demonstrate the basic integral formula. 

\subsubsection{The Basic Integral Formula}

In this subsection we will calculate the following integral

\begin{equation}
\label{apendice1}
J_\alpha(z)=\int_R^\infty\frac{e^{izy}}{y^{\alpha+1}}dy,
\end{equation}

\noindent where $\alpha>0$, $R>0$, $z\in\R$  and considering $y=|\vec u|$ and $z=\hat u\cdot\vec z$. 
For the transformation $\xi=zy$ with $z>0$, the integral in (\ref{apendice1}) may be written as
\begin{equation}
\label{apendice2}
J_\alpha(z)=z^\alpha\int_{zR}^\infty\frac{e^{i\xi}}{\xi^{\alpha+1}}d\xi.
\end{equation}
Considering $\alpha$ as a positive non-integer real number and taking the power series development of $e^{i\xi}$ up to order $p$ where $p$ is the greater integer lower than $\alpha$, that is, $p<\alpha<p+1$, then we have
\begin{equation}
\label{apendice3}
e^{i\xi}=1+i\xi+\frac{(i\xi)^2}{2!}+\cdots+\frac{(i\xi)^p}{p!} +{\cal R}(i\xi).
\end{equation}
Defining
\begin{equation}
\label{apendice4}
{\cal P}(i\xi)=\sum_{k=0}^p\frac{(i\xi)^k}{k!},
\end{equation}
the rest function in equation (\ref{apendice3}) can be written as
\begin{equation}
\label{apendice5}
{\cal R}(i\xi)=e^{i\xi}-{\cal P}(i\xi)=\sum_{k=p+1}^\infty\frac{(i\xi)^k}{k!}.
\end{equation}
Finally, with the definitions given above we can rearrange the integral in (\ref{apendice1}) in the following way
\begin{equation}
\label{apendice6}
J_\alpha(z)=z^\alpha\int_{zR}^\infty\frac{{\cal P}(i\xi)}{\xi^{\alpha+1}}d\xi-z^\alpha\int_0^{zR}\frac{{\cal R}(i\xi)}{\xi^{\alpha+1}}d\xi+z^\alpha
\int_0^{\infty}\frac{{\cal R}(i\xi)}{\xi^{\alpha+1}}d\xi.
\end{equation}
The first and the second integrals in (\ref{apendice6}) can be calculated as:
\begin{equation}
\label{apendice7}
\begin{array}{l}
\displaystyle\int_{zR}^\infty\frac{{\cal P}(i\xi)}{\xi^{\alpha+1}}d\xi =\sum_{k=0}^p\frac{i^k}{k!}\frac{(zR)^{k-\alpha}}{\alpha-k}\;\;(p<\alpha),\\
\displaystyle\int_0^{zR}\frac{{\cal R}(i\xi)}{\xi^{\alpha+1}}d\xi=
\sum_{k=p+1}^\infty\frac{i^k}{k!}\frac{(zR)^{k-\alpha}}{k-\alpha}\;\;(\alpha<p+1).
\end{array}
\end{equation}
The third integral in equation (\ref{apendice6}) may be put in the following form~\cite{levy4}:
\begin{equation}
\label{apendice8}
\int_0^{\infty}\frac{{\cal R}(i\xi)}{\xi^{\alpha+1}}d\xi
=i^{-\alpha}\frac{\Gamma(-\alpha)}{\Gamma(p+1-\alpha)}\int_0^\infty\frac{e^{-y}}{y^{\alpha-p}}dy=i^{-\alpha}\Gamma(-\alpha).
\end{equation}
Taking into account that
\[
i^{-\alpha}=e^{-i\alpha\pi/2}\;\;{\rm and}\;\; \Gamma(\alpha+1)\Gamma(-\alpha)=-\frac{\pi}{\sin(\pi\alpha)},
\]
the integral in (\ref{apendice8}) becomes
\begin{equation}
\label{apendice9}
\int_0^{\infty}\frac{{\cal R}(i\xi)}{\xi^{\alpha+1}}d\xi
=-\frac{\pi}{\Gamma(\alpha+1)}\frac{\cos\left(\alpha\pi/2\right)-i\sin\left(\alpha\pi/2\right)}{\sin(\alpha\pi)}.
\end{equation}
Inserting the integrals in (\ref{apendice7}) and (\ref{apendice9}) into equation (\ref{apendice6}) we get
\begin{equation}
\label{apendice10}
J_\alpha(z)=\sum_{k=0}^\infty\frac{R^{k-\alpha}}{k!|\alpha-k|}(iz)^k-\frac{\pi}{\Gamma(\alpha+1)}\frac{\cos(\alpha\pi/2)-i\sin(\alpha\pi/2)}{\sin(\alpha\pi)}z^\alpha.
\end{equation}

The equation (\ref{apendice10}) has been obtained considering $z>0$. In order to obtain the value of $J_\alpha(z)$ in (\ref{apendice1}) for $z<0$ one needs to consider that  
\[
J_\alpha(z)=J^{\rm Re}_\alpha(z)+iJ^{\rm Im}_\alpha(z),
\]
is a complex function with real and imaginary part, which are respectively even and odd functions, that is,
\begin{equation}
\label{apendice10a}
J^{\rm Re}_\alpha(-z)=J^{\rm Re}_\alpha(z);\;\;
J^{\rm Im}_\alpha(-z)=-J^{\rm Im}_\alpha(z).
\end{equation}
Thus for $z>0$:
\[
\begin{array}{l}
\displaystyle J^{\rm Re}_\alpha(z)=\sum_{k\;{\rm is\; even}}\frac{R^{k-\alpha}}{k!|\alpha-k|}i^kz^k-\frac{\pi}{\Gamma(\alpha+1)}\frac{\cos(\alpha\pi/2)}{\sin(\alpha\pi)}z^\alpha,\\
\displaystyle J^{\rm Im}_\alpha(z)=\sum_{k\;{\rm is\; odd}}\frac{R^{k-\alpha}}{k!|\alpha-k|}i^{k-1}z^k+\frac{\pi}{\Gamma(\alpha+1)}\frac{\sin(\alpha\pi/2)}{\sin(\alpha\pi)}z^\alpha.
\end{array}
\]
Therefore, for $z<0$ the even and odd summations above keep the same expression that for $z>0$, on the other hand $z^\alpha$ must be replaced by $|z|^\alpha$ in the real part $J^{\rm Re}_\alpha(z)$ and by $(z/|z|)|z|^\alpha$ in the imaginary part $J^{\rm Im}_\alpha(z)$. Thence, we can write for any real value of $z$:
\begin{equation}
\label{apendice11}
\displaystyle J_\alpha(z)=\sum_{k}^\infty\frac{R^{k-\alpha}}{k!|\alpha-k|}i^kz^k-\frac{\pi}{\Gamma(\alpha+1)}\frac{\cos(\alpha\pi/2)-i\sin(\alpha\pi/2)}{\sin(\alpha\pi)}|z|^\alpha.\\
\end{equation}	
Thus  we have showed that for non-integer positive $\alpha$ the integral $J_\alpha(z)$ can be written as a sum of a non analytical term plus a integer power series (see Ref.~\cite{levy4}).

Now, let consider the integral $J_\alpha(z)$ in (\ref{apendice1}) when $\alpha>0$ is an integer and $z>0$. Then, we integrate $\alpha$ times by parts to obtain
\begin{equation}
\label{apendice12}
J_\alpha(z)=\frac{e^{iRz}}{\alpha !R^\alpha}\sum_{k=0}^{\alpha-1}(\alpha-k-1)!(iz)^k+\frac{(iz)^k}{\alpha !} \int_R^\infty\frac{e^{iyz}}{y}dy.
\end{equation}
The integral in the right side of the equation (\ref{apendice12}) may be expanded in the following series
\begin{equation}
\label{apendice13}
\int_R^\infty\frac{e^{iyz}}{y}dy=-\gamma-\ln R+\frac{\pi}{2}i-\ln z+\sum_{k=1}^\infty\frac{R^k}{kk!}(iz)^k.
\end{equation}
Taking into account the parity properties (\ref{apendice10a}) of the real and imaginary parts of $J_\alpha(z)$ we can certainly write
\begin{equation}
\label{apendice14}
J_\alpha(z)=g(z)+f_\alpha(z)-\frac{(iz)^\alpha}{\alpha!}\ln|z|,
\end{equation}   
for all $z\in\R$, where $g(z)$ is a power series and $f_\alpha(z)$ is a polynomial function of order $\alpha$ for $z>0$. For instance, when $\alpha=1$ and $\alpha=2$ we have respectively
\[
f_1(z)=-\frac{\pi}{2}|z|+iz(1-\gamma-\ln R),
\]
\[
f_2(z)=-i\frac{\pi}{4}z|z|-z^2\left(\frac{3}{4}-\frac{\gamma}{2}-\frac{\ln R}{2}\right).
\]  

\subsubsection{The Canonical Characteristic Function}

The characteristic function of a random vector $\vec U$ with distribution  given  in Eq.~(\ref{equ2}) can be written as

\begin{equation}
{\psi_{\vec{U}} \left(\vec{z}\right)=\displaystyle \int _{\left|\vec{u}\right|\le u_c}e^{i\vec{u}\cdot \vec{z}}
f\left(\vec{u}\right)d^{d} \vec{u}
+\int_{\hat{u}\in S_{d} }\left[C\left(\hat{u}\right)\int _{u_c}^{\infty }
\frac{e^{i\left|\vec{u}\right|\hat{u}\cdot \vec{z}} }{\left|\vec{u}\right|^{1+\alpha } } d\left|\vec{u}\right| \right] dS_{d} }.
\label{equacao2}
\end{equation}

Taking into account the results of the previous subsection, the term inside the bracket in the right-hand side of Eq.~(\ref{equacao2}) can be written as:
\begin{equation}
\label{equacao3}
\int _{u_c}^{\infty }\frac{e^{i\left|\vec{u}\right|\hat{u}\cdot \vec{z}} }{\left|\vec{u}\right|^{1+\alpha } }
d\left|\vec{u}\right|=T_\alpha\left(\vec{z}\cdot\vec{u}\right)+g\left(\vec{z}\cdot \vec{u}\right),
\end{equation}
where $g\left(\vec{z}\cdot \vec{u}\right)$ is an integer power series on $\vec{z}\cdot \vec{u}$.
For non-integer values of $\alpha$ we have (see Eq.~\ref{apendice11}):
\begin{equation}
T_{\alpha }\left(\vec{z}\cdot\hat{u}\right)=-\frac{\pi\left|\vec{z}\cdot \hat{u}\right|^{\alpha }}{\Gamma \left(\alpha+1\right)}
\:\frac{\cos\left(\alpha\pi/2 \right)
-i\sin\left(\alpha\pi/2\right)(\vec{z}\cdot\hat{u})/\left|\vec{z}\cdot \hat{u}\right|}{\sin\left(\alpha \pi \right)},
\end{equation}
and for $\alpha$ integer (see Eq.~\ref{apendice14}):
\begin{equation}
T_{\alpha } \left(\vec{z}\cdot \vec{u}\right)=-\frac{i^\alpha}{\alpha!}\left(\vec{z}\cdot\vec{u}\right)^\alpha\ln\left|\vec{z}\cdot\vec{u}\right|+f_\alpha(\vec z\cdot\vec u).
\end{equation} 
For $\alpha\in\N$ it suffices to know the expressions for $\alpha=1$ and $\alpha=2$, which are respectively: 
\begin{equation}
T_{1} \left(\vec{z}\cdot \hat{u}\right)=-\frac{\pi }{2} \left|\vec{z}\cdot \hat{u}\right|
+i\vec{z}\cdot \hat{u}\left(1-\gamma -\ln u_c\right)-i\vec{z}\cdot \hat{u}\ln \left|\vec{z}\cdot \hat{u}\right|,
\end{equation}
\begin{equation}
T_{2} \left(\vec{z}\cdot \hat{u}\right)=-i\frac{\pi }{4} \vec{z}\cdot \hat{u}\left|\vec{z}\cdot \hat{u}\right|
-\frac{\left(\vec{z}\cdot\hat{u}\right)^{2}}{2}\left[\left(\frac{3}{2}-\gamma-\ln u_c\right)
+\ln\left|\vec{z}\cdot \hat{u}\right|\right].
\end{equation}

Now we proceed by noting that Eq.~(\ref{equacao2}) can be rewritten as:
\begin{equation}
\psi_{\vec{U}}\left(\vec{z}\right)=F(\vec z)+I_\alpha(\vec z),
\label{equacao4}
\end{equation}
where $F(\vec z)$ is an integer power series in $z$:
\begin{equation}
F(\vec z)=\int _{\left|\vec{u}\right|\le u_c}e^{i\vec{u}\cdot \vec{z}}f\left(\vec{u}\right)d^{d} \vec{u}
+\int _{\hat{u}\in S_{d} }C\left(\hat{u}\right)g\left(\vec{z}\cdot \hat{u}\right) dS_{d}
\label{equacao4b}
\end{equation}
and $I_\alpha$ is given for $\alpha \notin {\rm N}$ by: 
\begin{equation}
I_{\alpha } \left(\vec{z}\right)={\rm \; }-\frac{\pi }{\Gamma \left(\alpha +1\right)}
\frac{\cos \left({\alpha \pi  \mathord{\left/{\vphantom{\alpha \pi  2}}\right.\kern-\nulldelimiterspace} 2} \right)
A_{\alpha } \left(\hat{z}\right)-i\sin\left({\alpha \pi  \mathord{\left/{\vphantom{\alpha \pi  2}}\right.\kern-\nulldelimiterspace} 2} \right)
B_{\alpha } \left(\hat{z}\right)}{\sin\left(\alpha \pi \right)} \left|\vec{z}\right|^{\alpha },
\end{equation}
and
\begin{eqnarray}
I_{1}\left(\vec{z}\right) & = & \left[-\frac{\pi }{2} A_{1} \left(\hat{z}\right)+i\left[B_{1} \left(\hat{z}\right)
\left(1-\gamma -\ln u_c-\ln \left|\vec{z}\right|\right)-D_{1} \left(\hat{z}\right)\right]\right]\left|\vec{z}\right|,
\nonumber\\
I_{2}\left(\vec{z}\right) & = & -\left[A_{2} \left(\hat{z}\right)\left(\frac{3}{2} -\gamma -\ln u_c-\ln \left|\vec{z}\right|\right)
-D_{2} \left(\hat{z}\right)+i\frac{\pi }{2} B_{2} \left(\hat{z}\right)\right]\frac{\left|\vec{z}\right|^{2}}{2},
\nonumber\\
&
\end{eqnarray}
with $\hat z\equiv\vec{z}/|z|$ and the following definitions:
\begin{eqnarray}
\label{equacao5}
A_{\alpha }\left(\hat{z}\right) & = & \int _{S_{d} }C\left(\hat{u}\right)\left|\hat{z}\cdot \hat{u}\right|^{\alpha } dS_{d},
\nonumber\\
B_{\alpha }\left(\hat{z}\right) & = &\int _{S_{d} }C\left(\hat{u}\right)
\frac{\hat{z}\cdot \hat{u}}{\left|\hat{z}\cdot \hat{u}\right|} \left|\hat{z}\cdot \hat{u}\right|^{\alpha } dS_{d}, 
\nonumber\\
D_{\alpha }\left(\hat{z}\right) & = & \int _{S_{d} }C\left(\hat{u}\right)\left(\hat{z}\cdot \hat{u}\right)^{\alpha }
\ln\left|\hat{z}\cdot \hat{u}\right|dS_{d}.
\end{eqnarray} 

With the expressions for $I_\alpha(\vec z)$ and $F(\vec z)$, we have that the characteristic function $\psi_{\vec{U}}\left(\vec{z}\right)$ in Eq.~(\ref{equacao4}) is given by the sum of a power series in $\vec{z}$ and one non-analytical term from $I_\alpha(\vec z)$.
Taking into account the following series expansion:   
$$\ln \psi _{\vec{U}} \left(\vec{z}\right)=\ln \left(1+h\left(\vec{z}\right)\right)=h\left(\vec{z}\right)-\frac{1}{2} h^{2} \left(\vec{z}\right)+\cdots,$$
it is straightforward to show that the characteristic function  $\psi_{\vec{U}}\left(\vec{z}\right)$ can be written as (taking into account the different values of the Levy exponent $\alpha$):

\begin{equation}
\psi _{\vec{U}}\left(\vec{z}\right)=\exp\left(-|\vec{z}|^\alpha\left[\Xi_{\alpha}(\vec{z})+\Omega(\vec{z})\right]\right),
\label{equ3}
\end{equation}
for $\alpha\leq 1$,
\begin{equation}
\psi _{\vec{U}}\left(\vec{z}\right)=\exp\left(i\vec{z}\cdot\langle\vec U\rangle-|\vec{z}|^\alpha\left[\Xi_{\alpha}(\vec{z})+\Omega(\vec{z})\right]\right),
\label{equ4}
\end{equation}
for $1<\alpha\leq 2$ and 
\begin{equation}
\psi _{\vec{U}}\left(\vec{z}\right)=\exp\left(i\vec{z}\cdot\langle\vec U\rangle-|\vec{z}|^2\left[\Xi_{\alpha}(\vec{z})+\Omega(\vec{z})\right]\right),
\label{equ4a}
\end{equation} 
for $\alpha> 2$. Where $\Omega\left(\vec{z}\right)$ is a continuous function in a neighborhood of $\vec{z}=0$ such that
\begin{equation}
\label{equ5}
\displaystyle
\lim_{\vec{z}\rightarrow 0}
\Omega\left(\vec{z}\right)=\Omega(0)=0.
\end{equation}
The equations (\ref{equ3}), (\ref{equ4}) or (\ref{equ4a}) are the canonical form for the characteristic function of a power law distribution of random vectors.

The analytical form of the function $\Xi_{\alpha}(\vec{z})$ relies on which interval one has for $\alpha$ and  the expressions are presented below according to each interval for the quantity $\alpha$:  
\begin{equation}
\label{equ6}
\Xi_{\alpha}(\vec{z})=
\frac{\pi }{\Gamma \left(\alpha +1\right)} \frac{\cos \left({\alpha \pi  \mathord{\left/{\vphantom{\alpha \pi  2}}\right.\kern
-\nulldelimiterspace} 2} \right)A_{\alpha } \left(\hat{z}\right)-i\sin\left({\alpha \pi  \mathord{\left/{\vphantom{\alpha \pi  2}}\right.\kern-\nulldelimiterspace} 2} \right)
B_{\alpha } \left(\hat{z}\right)}{\sin\left(\alpha \pi \right)}
\end{equation} 
for $0<\alpha<2$ ($\alpha\neq 1$),
\begin{eqnarray}
\label{equ7}
\lefteqn{\Xi_{1}(\vec{z})=\frac{\pi }{2} A_{1} \left(\hat{z}\right)}
\nonumber\\
 & & -i\left[\hat z\cdot\langle\vec U\rangle_{u_c}+
B_{1}\left(\hat{z}\right)\left(1-\gamma -\ln u_c\right)+B_{1} \left(\hat{z}\right)\ln \left|\vec{z}\right|-D_{1} \left(\hat{z}\right)
\right].
\end{eqnarray}
for $\alpha=1$,
\begin{eqnarray}
\label{equ8}
\lefteqn{\Xi_2(\vec{z})=\hat{z}\cdot\bar{M}_{u_c}\cdot\hat{z}}
\nonumber\\
 & & +\left[A_2(\hat z)\left(\frac{3}{2}-\gamma-\ln u_c\right)-D_2(\hat z)
-A_2(\hat z)\ln|\vec z|+i\frac{\pi}{2}B_2(\hat z)\right].
\end{eqnarray} 
for $\alpha=2$, and 
\begin{equation}
\label{equ9}
\Xi_{\alpha}(\vec{z})=\frac{1}{2}\:\hat z\cdot\bar{M}\cdot\hat z,
\end{equation}
with $\hat z\equiv\vec{z}/|z|$ for $2<\alpha<\infty$.

The matrix $\bar{M}$ is the covariance matrix in Eq.~(\ref{equ1}),
$\langle\vec U\rangle_{u_c}$ is the average of $\vec U$ computed for $|\vec u|\leq u_c$
$$
\langle\vec U\rangle_{u_c}=\int_{|\vec u|\leq u_c}\vec u\rho_{\vec U}(\vec u)d^n\vec u,
$$
and $\bar M_{u_c}$ is similar to a covariance matrix in Eq.~(\ref{equ1}) but computed as:
\begin{equation}
\left(\bar M_{u_c}\right)_{ij}=\langle U_iU_j\rangle_{u_c}-\langle U_i\rangle\langle U_j\rangle,
\end{equation}
\begin{equation}
\langle U_iU_j\rangle_{u_c}\equiv\int_{\left|\vec{u}\right|\le u_c}u_{i} u_{j} \rho _{\vec{U}} \left(\vec{u}\right)d^{n}\vec{u}.
\end{equation}

\section{L\'evy-Khintchine Stable Distributions of the Resultant Force}
\label{LK}

In this section we obtain the distribution function of the resultant force $\vec F_{res}^N$ on the test particle, which is given by the vector sum of the $N$ forces $\vec{F}_k$ in the sequence, that is,
\begin{equation}
\vec{F}^N_{res} =\sum _{k=1}^{N}\vec{F}_{k}.
\end{equation}
For the approach we consider a sequence of $N$ identical and independent random forces $\vec{F}_1,\vec{F}_2,\ldots,\vec{F}_N$ acting on the test particle, with
$\vec{F}_k=\vec{U}$ for all $k=1,\ldots,N$ and the density distribution $\rho_{\vec{U}}\left(\vec{u}\right)$ is given by Eq.~(\ref{equ2}).
Taking each $\vec{F}_k$ of the sequence ($\vec{F}_1,\vec{F}_2,\ldots,\vec{F}_N$) as mutually independent and identical, the characteristic function of the resultant force is given by
\begin{equation}
\psi _{\vec{F}^{N}_{res} } \left(\vec{z}\right)=
\prod_{k=1}^N\psi_{\vec{F}_k}\left(\vec{z}\right)=
\psi _{\vec{U}} \left(\vec{z}\right)^{N}.
\label{equ11}
\end{equation}
Given a random vector variable $\vec{F}^{N}_{res}$, a real number $a_N$ and a $n$-dimensional vector $\vec{b}_N$, we define the renormalization of $\vec{F}^N_{res}$ by:
\begin{equation}
\label{equ12}
\bar{F}^{N}_{res} =a_{N} \vec{F}^{N}_{res} +\vec{b}_{N}.
\end{equation}
Its characteristic function is given by: 
\begin{equation}
\label{equ13}
\psi _{\bar{F}_{res}^N } \left(\vec{z}\right)=\exp\left(\vec{b}_{N} \cdot \vec{z}\right)\psi _{\vec{U}}\left(a_{N} \vec{z}\right)^{N}.
\end{equation}
If the sequence of characteristic functions in Eq.~(\ref{equ13}) converges to a well defined continuous function:
\begin{equation}
\lim_{N\rightarrow\infty}\psi _{\bar{F}^{N}_{res} } \left(\vec{z}\right)=\Phi\left(\vec{z}\right)
\label{equ14}
\end{equation}
then, as a consequence of L\'evy Continuity Theorem, the sequence of density distributions associated to the renormalized forces ${\bar F^1_{res}}, {\bar F^2_{res}}.\ldots$ converges to a well defined density distribuition for $N\rightarrow\infty$.
We then say that the sequence of random resultant forces $\vec{F}^1_{res}, \vec{F}^2_{res},\ldots$ weakly converges in distribution.

The renormalization of each random resultant force in the sequence $\vec{F}^1_{res}$, $\vec{F}^2_{res},\ldots$ is chosen in a way that the renormalized sequence $\bar{F}^1_{res},\bar{F}^2_{res},\ldots$ converges and is obtained according to the value of $\alpha$. The different possibilities for the limit characteristic function in (\ref{equ14}) and the respective renormalizations are obtained in the appendix \ref{appb}. The results are sumarized in Table~\ref{tab1}.

\begin{table}[!htbp]

\caption{\it The Logarithm of the Limit Characteristic Function $\Phi({\vec z})$ and their respective renormalizations. 
For all cases where the resultant force has a mean we have $<{\vec F}^N_{res}>=N<{\vec U}>$.
The vector ${\vec v}$ and the matrix $M$ are defined by formulas
(\ref{equacao16a}) and (\ref{equacao20a}) respectively given in Appendix~\ref{appb}.}
\begin{center}
\begin{tabular}{c}
\hline\hline
Non Singular Cases \\ \hline\hline
\\
\begin{tabular}{cc}
$[0<\alpha<1]$ & $[1<\alpha<2]$ \\
$a_N=1/N^{1/\alpha}$\;\;${\vec b}_N=0$ &
$a_N=1/N^{1/\alpha}$\;\;${\vec b}_N=-\left<{\vec F}^N_{res}\right>/N^{1/\alpha}$
\end{tabular}\\ 
$\displaystyle\ln\Phi({\vec z})=
-\left|\vec{z}\right|^{\alpha }\frac{\pi}{\Gamma(\alpha+1)}\:\frac{\cos(\alpha\pi/2)A_\alpha(\hat{z})-i\sin(\alpha\pi/2)B_\alpha(\hat{z})}{\sin(\alpha\pi)}$\\ \\
$[2<\alpha<\infty]$\\
\begin{tabular}{cc}
$a_N=1/N^{1/2}$\;${\vec b}_N=-\left<{\vec F}^N_{res}\right>/N^{1/2}$ &
$ \displaystyle\ln\Phi({\vec z})=-
\frac{1}{2}\:\vec z\cdot\bar{M}\cdot\vec z$
\end{tabular}
\\
\\\hline\hline \\ \hline\hline
Singular Cases \\ \hline\hline\\
$[\alpha=1]$\;\;$\displaystyle a_N=\frac{1}{N}\;\vec{b}_N=-\langle\vec U\rangle_{u_c}+\vec v\left(1-\gamma-\ln u_c\right)-\vec v\ln N$ \\
$\displaystyle\ln\Phi({\vec z})=
-|\vec{z}|\left\{\frac{\pi}{2}A_1(\hat z)
-i\left[\hat z\cdot\vec v\ln|\vec z|+D_1(\hat z)\right]\right\}$
\\ \\
$[\alpha=2]$\;\; $\displaystyle a_N=\frac{1}{\left(N\ln N\right)^{1/2}}\;
\vec{b}_N=-\frac{\langle\vec{F}^N_{res}\rangle}{\left(N\ln N\right)^{1/2}}$\\$\displaystyle\ln\Phi({\vec z})=
-\frac{1}{2}\:\vec z\cdot M \cdot\vec z$\\ \\\hline\hline
\end{tabular}
\end{center} 
\label{tab1}
\end{table}

The results shown in Table \ref{tab1} for $0<\alpha\leq 2$ ($\alpha\neq 1$), can all be put together in a canonical form that  clearly emerges as a generalization of the L\'evy-Khintchine stable distributions for random real variables. 
In analogy with scalar random variables,
we define a L\'evy-Khintchine stable distribuition for random vectors as the density distribution for which the corresponding characteristic function can be written as:
\begin{equation}
\label{equ15}
\Phi(\vec{z})=\exp\left(
-\frac{\lambda A_\alpha(\hat{z})|\vec{z}|^\alpha}{\Gamma(1+\alpha)}\left[1-i\beta_\alpha(\hat{z})
\tan\left(\frac{\alpha\pi}{2}\right)\right]\right),
\end{equation}
with $\beta_{\alpha}(\hat{z})=B_\alpha(\hat{z})/A_\alpha(\hat{z})$ and
$A(\hat{z})$ and $B(\hat{z})$ defined in Eq.~(\ref{equacao5}) and $\lambda$ is any positive real number. We have used the trigonometric identity $\sin(\alpha\pi)=2\sin(\alpha\pi/2)\cos(\alpha\pi/2)$, factors out $\pi/\left(2\sin\left(\alpha\pi/2\right)\right)$, and incorporates in $\lambda$. 

The characteristic function written in the form given in Eq.~(\ref{equ15}) is directly related with the standard characteristic function of stable random vectors presented by Samorodtsky and Taqqu in reference \cite{taqqu}.
Some well known properties of the L\'evy-Khintchine stable distributions for random vectors are presented in Appendix \ref{appc}.

For $\alpha=1$ the characteristic function does not represent a true stable probability distribution, provided that the function $C(\hat u)$ defined in (\ref{equ2}) is not symmetric in the hypersphere $S_d$. In this case we have that $\vec v\neq 0$, $D_1(\hat z)\neq 0$ and the respective non symmetric probability distribution is called semi-stable. The semi-stable probability densities of real random variable was first introduced by Levy \cite{levy2}. 

However, if the function $C(\hat u)$ is symmetric, that is, $C(\hat u)=C(-\hat u)$, then $\vec v=0$ and $D_1(\hat z)=0$ and the corresponding characteristic function represents a symmetric stable probability distribution which constitutes a generalization of the Cauchy 
distribution of a real random variable to a $d$-dimensional random vector.

For $\alpha>2$ the renormalized sum of identical random vectors $\vec{U}$ converges to a Gaussian distribution,
independently of the specific statistical properties of the density distribution $\rho_{\vec{U}}(\vec{u})$. Moreover, the only property of $\vec{U}$
that is relevant in the $N\rightarrow\infty$ limit is its covariance structure given by $\bar{M}$.

Note that for $N\to\infty$ the random vector sequence $\bar F^N_{res}$ converges to a random vector denoted by $\vec S_\alpha$ with a probability density
given by a L\'evy stable distribution with characteristic function given in Table 1 for $0<\alpha<2$ ($\alpha\neq 1$). 

Also, from the Table 1, we can see that for $\alpha=1$ the random sequence $\bar F^N_{res}$ converges to a random vector denoted by $\vec S_1^{(i)}$($i=0,1$), and $i=0$ corresponds to a stable L\'evy distribution and $i=1$ corresponds to a semi-stable distribution.
For $\alpha\ge 2$ the probability distribution is a Gaussian distribution with associated random vector denoted by $\vec S^*_2$ for $\alpha=2$ and $\vec S_2$ for $\alpha>2$. 

It is worth to remark that both random vectors $\vec S_2^*$ and $\vec S_2$ are given by a Gaussian distribuition, but only for $\vec S_2$ the Gaussian distribution is defined by the covariance matrix ${\bar M}$ of the random force ${\vec U}$. Indeed, the covariance matrix of $\vec U$ can not be defined for $\alpha=2$ and the limit Gaussian in this case is defined by the matrix $M$ given in equation (\ref{equacao20a})

We finalize this section discussing the behaviour of the asymptotic random resultant force when the number of particles tends to infinity. For that, we take the limit $N\rightarrow\infty$ in the sequence (\ref{equ12}) to obtain:

\begin{equation}
\label{equ16}
\lim_{N\rightarrow\infty}\vec F^N_{res}=\lim_{N\rightarrow\infty}\left(\frac{1}{a_N}\right)\vec S_\alpha-\lim_{N\rightarrow\infty}\frac{\vec b_N}{a_N}.
\end{equation}

\noindent Equation (\ref{equ16}) may be decomposed for different ranges of $\alpha$ and one gets, assymptotically, for the random resultant force: 

\begin{equation}
\label{equ17}
\lim_{N\rightarrow\infty}\vec F^N_{res}=
\left\{\begin{array}{lc}
N^{1/\alpha}\vec S_\alpha & 0<\alpha<1\\ & \\
N\ln N\vec v+N\vec S_1^{(i)} & \alpha=1\;(i=0,1)\\ & \\
N\left<\vec U\right>+N^{1/\alpha}\vec S_\alpha & 1<\alpha<2\\ & \\
N\left<\vec U\right>+(N\ln N)^{1/2}\vec S^*_2 & \alpha=2\\ & \\
N\left<\vec U\right>+N^{1/2}\vec S_2 & 2<\alpha
\end{array}\right.
\end{equation}
In the next section we present the results for Eq.~(\ref{equ17}) in the context of some physical scenarios. 

\section{Results}
\label{prob}
\subsection {The Resultant Limit Force  for a Central Power Law Random Force}

We begin with the calculation of the distribution probability for the resultant force on the test particle due to $N$ point particles distributed in space according to a give spatial density of probability. 
The approach is based on associating  a random vector $\vec{R}$ to the relative position of the $N$ particles with respect to the test particle which is characterized by its probability distribution $\rho_{\vec{R}}(\vec{r})$ with $\vec{r}\in\R^d$, and $d=1,2,3$ standing for the spatial dimension.

The force $\vec{F}^N_i$ on the test particle due to particle $i$, $i=1,\ldots,N$ can be represented by a random vector transformation:

\begin{equation}
\label{extra1}
\vec F^N_i=\vec F^N(\vec R).
\end{equation} 

In this point one assumes that the $N$ random forces are statistically identical. The resultant force $\vec{F}^N_{res}$ is the vector sum of the $N$ forces $\vec{F}^N_i$, $i=1,\ldots,N$:

\begin{equation}
\label{equ18}
\vec{F}^N_{res}=\sum_{i=1}^N\vec{F}^N_i,
\end{equation}

\noindent and the renormalized force $\vec {F}^N_i$ is given by:

\begin{equation}
\label{equ19}
\vec F_i^N(\vec R)=a_N\,\vec U(\vec R),
\end{equation}

\noindent with $N$ large and $a_N$ is constant. It is usual in the approach for long range interacting systems to rescale the potencial energy with the system size, e.g., introducing the Kac factor~\cite{kac}, and for that one includes in the representation of the force the superscript $N$, or $\vec F_i^N$.

Considering that the spatial position of the particles are uncorrelated, which is always true if the inter-particle interaction is long-ranged~\cite{steiner},
i.e., that the position vectors of the particles are statisticaly independent, we can write the characteristic function of the resultant random force as:
\begin{equation}
\label{equ20}
\psi _{\vec{F}^N_{res}}(\vec{z})=\prod_{i=1}^N\psi_{\vec{F}^N_i}(\vec{z})=\psi_{(a_N\vec{U})}(\vec{z})^N=\psi_{\vec{U}}(a_N\vec{z})^N.
\end{equation} 
If the limit of Eq.~(\ref{equ20}) for $N\rightarrow\infty$ exists (keeping other thermodynamics variables constant) then its characteristic function, denoted by $\Phi(\vec{z})$, may be calculated according to the following limit:
\begin{equation}
\label{equ21}
\Phi(\vec{z})=\lim_{N\rightarrow\infty}\psi_{\vec{F}^N_{res}}(\vec{z})=\lim_{N\rightarrow\infty}\psi_{\vec{U}}(a_N\vec{z})^N.
\end{equation}

In order to proceed consider a force $F_i^N(\vec R)$ given by:
\begin{equation}
\label{equ22}
\vec{F}_i^N=\kappa_N\frac{\hat{R}}{|\vec{R}|^\delta},\;\;\hat{R}=\frac{\vec{R}}{|\vec{R}|},
\end{equation}   
with $\kappa_N$ and $\delta$ being constant real numbers and $\delta\ge 0$. For such a force (Eq.~(\ref{equ22})) we should make use of Eq.~(\ref{equ21}) to determine the probability distribution of the resultant force in the limit $N\rightarrow\infty$. To carry out the calculation, we define the renormalized variable $\vec{U}=\hat{R}/|\vec{R}|^\delta$ with probability distribution denoted as
$\rho_{\vec{U}}(\vec{u})$ and $\vec{u}\equiv\hat{r}/|\vec{r}|^\delta$.
Denoting by $\rho_{\vec{R}}$ the spatial distribution of particles, i.e., the distribution of $\vec{R}$, we have:
\begin{equation}
\label{equ23}
\rho_{\vec{U}}(\vec{u})=\rho_{\vec{R}} (\vec{r})\left\|\frac{\partial\vec{r}}{\partial\vec{u}}\right\|=
\frac{1}{\delta}\,\frac{\rho_{\vec{R}}(\hat u/|\vec{u}|^{1/\delta})}{|\vec u|^{d+d/\delta}}.
\end{equation}

To apply the results from the previous section we have to determine the tail of the distribution, i.e., the asymptotic behaviour of 
$\rho_{\vec{U}}(\vec{u})$
for $|\vec u|\rightarrow\infty$. From the definition of $\vec{u}$ and Eq.~(\ref{equ23}) we have:
\begin{equation}
\label{equ24}
\lim_{|\vec u|\to \infty }\rho_{\vec{U}}(\vec{u})=\frac{1}{\delta|\vec u|^{d+d/\delta}}\lim_{|\vec r|\to 0}\rho_{\vec{R}}(\vec{r}).
\end{equation}
The assymptotic behaviour of $\rho _{\vec{U}}(\vec u)$ is thus determined by the short distance behaviour of $\rho _{\vec{R}}(\vec{r})$.

For such systems described by this sort of force one can assume that exists a constant $r_c>0$ (possibly $r_c<<1$) such that the density distribution
$\rho_{\vec{R}}(\vec{r})$ for $|\vec r|<r_c$ is given by:
\begin{equation}
\label{equ25}
\rho _{\vec{R}} \left(\vec{r}\right)=g\left(\hat{r}\right)|\vec r|^{\nu } ;{\rm \; \; }\nu >-d.
\end{equation}
From Eq.~(\ref{equ23}) we obtain 
\begin{equation}
\label{equ26}
\rho_{\vec{U}}(\vec{u})=\frac{g(\hat{r})|\vec r|^\nu}{\delta|\vec u|^{d+d/\delta}}
=\frac{g(\hat{u})(1/|\vec u|^{1/\delta})^\nu}{\delta|\vec u|^{d+d/\delta}}=
\frac{g(\hat{u})}{\delta}\frac{1}{|\vec u|^{d+d/\delta+\nu/\delta}},
\end{equation}
for $|\vec r|<r_c$ or equivalently $|\vec u|>u_c=1/r_c^\delta$. Thus the density distribution $\rho_{\vec{U}}(\vec{u})$ can be written as:
\begin{equation}
\label{equ27}
\rho_{\vec{U}}(\vec{u})=\left\{
\begin{array}{l}
f(\vec{u}),\hspace{3mm}{\rm if}\hspace{2mm}|\vec u|\le u_{c},
\\
\\
\displaystyle{\frac{C(\hat{u})}{|\vec u|^{d+\alpha}},\hspace{3mm}{\rm if}\hspace{2mm}|\vec u|>u_{c}},
\end{array}
\right.
\end{equation} 
with
\begin{equation}
\alpha=(d+\nu)/\delta,
\label{equ28}
\end{equation}
and
\begin{equation}
C(\hat{u})=g(\hat{u})/\delta.
\label{equ29}
\end{equation}
Note that this distribution has the same form as in Eq.~(\ref{equ2}) and its respective characteristic functions
are determided in Eqs.~(\ref{equ3}--\ref{equ9}). 

To obtain the distribution of the resultant random force $\vec{F}^N_{res}$, we renormalize the force $\vec{F}_i^N$ by a suitably chosen of the force constant $\kappa_N$ or (and) the characteristic length $L_N>0$:
\begin{equation}
\label{equ30}
\vec{F}^N_i=\kappa_N\frac{\hat R}{{|L_N\vec{R}|}^{\delta}}=\frac{\kappa_N}{L_N^\delta}\frac{\hat R}{{|\vec R|}^\delta}=a_N\vec{U},
\hspace{5mm}a_N=\frac{\kappa_N}{{L_N}^\delta},
\end{equation}
in such a way that the limit defined in Eq.~(\ref{equ21}) exists.

\subsubsection{The Vlasov Limit and the Fluctuation Limit}

In this section we calculate the limit of the resultant force for $N\rightarrow\infty$ considering a random vector $\vec U$ with density distribution given in Eq.~(\ref{equ27}) and L\'evy exponent $\alpha>0$. 
For this purposes we insert the respective characteristic functions of $\vec U$ into the limit defined in Eq.~(\ref{equ21}) to obtain:

\begin{equation}
\label{equ31}
\ln\Phi(\vec{z})=
\lim_{N\rightarrow\infty}\left\{
\begin{array}{l}
\hspace*{-2mm}\displaystyle -N|a_N|^\alpha|\vec{z}|^{\alpha}
\left[\Xi_{\alpha}(q\hat z)+\Omega(a_N\vec z)\right]\;(0<\alpha<1)
\\ \\
\hspace*{-2mm}\displaystyle -N|a_N||\vec z|\left[\Xi_1(a_N\vec z)+\Omega\left(a_N\vec z\right)\right]\;(\alpha=1) 
\\ \\
\hspace*{-2mm}\displaystyle iN|a_N|\vec{z}\cdot\langle q\vec U\rangle-N|a_N|^\alpha|\vec{z}|^{\alpha}
\left[\Xi_{\alpha}(q\hat z)+\Omega(a_N\vec z)\right]\;(1<\alpha<2)
\\ \\
\hspace*{-2mm}\displaystyle iN|a_N|\vec{z}\cdot
\langle q\vec{U}\rangle
-N|a_N|^2 |\vec z|^2 \left[\Xi_2(a_N\vec z)+\Omega(a_N\vec{z})\right]\;(\alpha=2)
\\ \\
\hspace*{-2mm}\displaystyle iN|a_N|\vec{z}\cdot
\langle q\vec{U}\rangle-N|a_N|^2|\vec{z}|^2\left[\Xi_{\alpha}(q\hat z)+\Omega\left(a_N\vec{z}\right)\right]\;(2<\alpha),
\end{array}
\right.
\end{equation}

\vspace*{2mm}
\noindent where $q=a_N/|a_N|={\rm sgn}(\kappa_N)$.

The expressions for $\Xi_\alpha(\hat z)$ with $\alpha\neq 1,2$, given in Eqs.~(\ref{equ6}) and (\ref{equ9}), may be written respectively as: 
\begin{equation}
\Xi_\alpha(q\hat z)=\frac{G_\alpha(q\hat z)}{\Gamma(\alpha+1)}\;(0<\alpha<2),\;\;\Xi_\alpha(q\hat z)=\frac{\hat z\cdot\bar M\cdot\hat z}{2}\;(2<\alpha),
\end{equation}
with the following definition:
\begin{equation}
\label{equ32}
G_\alpha(q\hat z)=\Gamma(\alpha+1)\Xi_\alpha(q\hat z)=
\pi\frac{\cos(\alpha\pi/2)A_\alpha(\hat{z})-i\sin(\alpha\pi/2)B_\alpha(q\hat{z})}{\sin(\alpha\pi)}.
\end{equation}

The functions $\Xi_1(\vec z)$ and $\Xi_2(\vec z)$, respectively defined in Eqs.~(\ref{equ7}) and (\ref{equ8}), 
allows us to write:
\begin{eqnarray}
\Xi_1(a_N\vec z)&=&-i\left[B_1(q\hat z)\ln|a_N\vec z|+R_1(q\hat{z})\right]+\frac{\pi}{2}A_1(q\hat z),
\nonumber\\
\Xi_2(a_N\vec z)&=&-A_2(\hat z)\ln|a_N\vec{z}|+R_2(q\hat z).
\nonumber
\end{eqnarray}

The possible renormalizations of $a_N$ are given in order to assure a well defined limit in Eq.~(\ref{equ31}) and are sumarized in Table \ref{tab2}.

\begin{table}[h!tb]

\caption{\it The Renormalizations of $a_N$. $K$ is an arbitrary positive constant}
\begin{center}
\begin{tabular}{c}
\begin{tabular}{ccc}
\hline\hline
Fluctuation ($0<\alpha<1$) & Singular ($\alpha=1$) & Vlasov ($1<\alpha<\infty$)   
\\ \hline\hline \\
$N|a_N|^\alpha=K$ & $-N|a_N|\ln |a_N|=K$ & $N|a_N|=K$  \\ \\
$\displaystyle |a_N|=\left(\frac{K}{N}\right)^{1/\alpha}$ & $\displaystyle |a_N|=h(K/N)$ &
$\displaystyle |a_N|=\frac{K}{N}$ \\ \\
$N>0$ & $\displaystyle N\geq\frac{2}{\ln 2}K$ & 
$N>0$ \\ \\
\hline\hline
\end{tabular}
\end{tabular}
\end{center}
\label{tab2}
\end{table}
\noindent For all renormalization cases presented in the Table \ref{tab2}  we have that
$$
\lim_{N\rightarrow\infty}a_N=0\;\;\Rightarrow\;\;\lim_{N\rightarrow\infty}\Omega(a_N\vec z)=0.
$$ 
For $\alpha=1$ and $\alpha=2$ we have respectively:
\begin{eqnarray}
&\displaystyle \lim_{N\rightarrow\infty}\left\{N|a_N|\Xi_1(a_N\vec z)\right\}=\lim_{N\rightarrow\infty}\left\{-iN|a_N|\ln|a_N|B_1(q\hat z)
+N|a_N|\frac{\pi}{2}A_1(q\hat z)\right\},&
\nonumber\\
&\displaystyle \lim_{N\rightarrow\infty}\left\{N|a_N|^2\Xi_2(a_N\vec z)\right\}=\lim_{N\rightarrow\infty}\left\{-N|a_N|^2\ln|a_N|\right\}A_2(\hat z).&
\nonumber
\end{eqnarray}
Then, from Eq.~(\ref{equ31}) we can obtain:
\begin{equation}
\label{equ33}
\ln\Phi(\vec{z})=\lim_{N\rightarrow\infty}
\left\{
\begin{array}{l}
\hspace*{-2mm}\displaystyle -\frac{\sigma_N^\alpha|\vec{z}|^\alpha}{\Gamma(\alpha +1)}G_\alpha(q\hat z)\;\;
\sigma_N=K^{1/\alpha} \;\;(0<\alpha<1)
\\ \\
\hspace*{-2mm}\displaystyle i\vec z\cdot(qK\vec v)-\sigma_N \frac{\pi A_1(\hat z)}{2}|\vec z|\;\;\sigma_N=Nh(K/N)\;\;(\alpha=1)
\\ \\
\hspace*{-2mm}\displaystyle i\vec{z}\cdot\langle qK\vec U
\rangle-\frac {\sigma_N^\alpha|\vec{z}|^\alpha 
G_\alpha(q\hat z)}{\Gamma(\alpha +1)} 
\;\; \sigma_N=KN^{(1-\alpha)/\alpha}
\;\;(1<\alpha<2)
\\ \\
\hspace*{-2mm}\displaystyle i\vec{z}\cdot\langle qK\vec U\rangle-\frac{\sigma_N^2}{2}\vec{z}\cdot M\cdot\vec{z}
\;\; \sigma_N=K\left(N^{-1}\ln N\right)^{1/2} \;\;(\alpha=2) 
\\ \\
\hspace*{-2mm}\displaystyle i\vec{z}\cdot
\langle{qK\vec U}\rangle-\frac{\sigma_N^2}{2}\vec{z}\cdot\bar{M}\cdot\vec{z}
\;\;\sigma_N=KN^{-1/2}\;\;(2<\alpha<\infty)\; ,
\end{array}\right. 
\end{equation}

\vspace*{2mm}
\noindent The vector ${\vec v}$ and the matrix $M$ are defined respectively by formulas (\ref{equacao16a}) and (\ref{equacao20a}) given in Appendix~\ref{appb} 

Using the L\'evy's Continuity Theorem, the limit renormalization process above for the characateristic functions can be presented in terms of the random resultant forces as follows:
\begin{equation}
\label{equ34}
\lim_{N\rightarrow\infty}{\vec F}_{res}^N=\lim_{N\rightarrow\infty}
\left\{
\begin{array}{l}
\hspace*{-2mm}\displaystyle \sigma_N{\vec S}_\alpha\;\;
\sigma_N=K^{1/\alpha} \;\;(0<\alpha<1)
\\ \\
\hspace*{-2mm}\displaystyle qK\vec v+\sigma_N{\vec S}^0_1\;\;
\sigma_N=Nh(K/N)\;\; (\alpha=1)
\\ \\
\hspace*{-2mm}\displaystyle \langle qK\vec U\rangle+\sigma_N{\vec S}_\alpha\;\;
\sigma_N=KN^{(1-\alpha)/\alpha}\;\;(1<\alpha<2)
\\ \\
\hspace*{-2mm}\displaystyle \langle qK\vec U\rangle+\sigma_N{\vec S}_2^*\;\;
\sigma_N=K\left(N^{-1}\ln N\right)^{1/2}\;\;(\alpha=2) 
\\ \\
\hspace*{-2mm}\displaystyle \langle qK\vec U\rangle+\sigma_N{\vec S}_2\;\;
\sigma_N=KN^{-1/2}\;\;(2<\alpha<\infty),
\end{array}\right. 
\end{equation}
where the L\'evy stable random vectors ${\vec S}_\alpha$ are defined at the end of section 2.
 
From the assymptoptic expressions for the global dispersion parameter $\sigma_N$ in Eqs.~(\ref{equ33})~or~(\ref{equ34}), we can conclude that 
$$\lim_{N\rightarrow\infty}\sigma_N=0
\;\;\Rightarrow\;\;\ln\Phi(\vec z)=i<qK\vec U>,$$
for all L\'evy exponent $\alpha>1$. This result suggest that the resultant force converges to the mean resultant force:

\begin{equation}
\label{equ35}
\vec F_{mf}\equiv\lim_{N\to\infty}
\langle\vec F_{res}^N\rangle=\langle qK\vec U\rangle=\lim_{N\rightarrow \infty}{\vec F}_{res}^N,
\end{equation}

\noindent and {\it mf} stands for {\it mean-field} as this limit corresponds to the mean-field limit in statistical mechanics.
Indeed, the renormalization for $\alpha> 1$ is guided by the requirement that the resultant force have a well defined mean, which is independent of the particle number $N$. This limit we named as the {\it{Vlasov Limit}}.  
For $N\rightarrow\infty$ the assymptotic resultant force in the {\it{Vlasov Limit}} can be written as
\begin{equation}
\vec{F}^N_{res}=\vec{F}_{mf}+\delta\vec{F}^N,
\end{equation}
where $\delta\vec{F}^N=\vec{F}^N_{res}-\vec{F}_{mf}$ is the fluctuation of the resultant force around its mean and
its statistic is described by a stable L\'evy random vector $\vec S_\alpha$ for $\alpha>1$. 

The fluctuation about the mean force has a global dispersion parameter given by $\sigma_N$ and goes to zero whenever 
$N\rightarrow\infty$. For $\alpha\ge 2$ the statistical distribution of $\delta\vec F^N$ tends to a Gaussian distribution and, when $\alpha> 2$,
$\sigma_N$ contributes to the random resultant force as a multiplicative factor of the standard deviation for each direction of $\vec{F}^N_{res}$. 

For the singular case $\alpha=1$ the random force $\vec F^N_i$ has no finite mean. However, from Eq.~(\ref{equ33}) or Eq.~(\ref{equ34}) and the fact that $\lim_{N\rightarrow\infty}\sigma_N=0$ we can conclude that
\begin{equation}
\ln\Phi(\vec z)=i\vec z\cdot (qK\vec v)\;\;\iff\;\;\lim_{N\rightarrow\infty}\vec F^N_{res}=qK\vec v,
\end{equation}
and, in a similar way as in the {\it{Vlasov Limit}}, the resultant force converges to a well defined vector for $N\rightarrow\infty$. In this case we can also define a Fluctuation Force about this vector $qK\vec v$ with statistic described by the stable symmetric Cauchy vector ${\vec S}^0_1$.

To conclude the analysis we note that for 
$0<\alpha<1$ the random forces $F^N_i$ have no finite mean, and, differently to the singular case $\alpha=1$, no possible renormalization can be defined in order to assure that the resultant force converges to a well defined vector for $N\rightarrow\infty$.

As a consequence, the resultant force 
is purely due to fluctuations and a {\it{Vlasov Limit}} as made previously can not be defined. The interval $0<\alpha<1$ we named as {\it{Fluctuation Limit}}.
In this case a well defined limit for the density distribution of the resultant force is still possible, provided the parameter $a_N$  is renormalized as shown in the Table 2. Indeed, from Eq.~(\ref{equ33}) it is straightfoward to see that:

\begin{equation}
\label{equ36}
\ln\Phi(\vec{z})=-\frac{K|\vec{z}|^\alpha G_\alpha(q\hat z)}{\Gamma(\alpha +1)}.
\end{equation}

The statement that the random resultant force is a pure fluctuating force can be expressed as:

\begin{equation}
\label{equ37}
\lim_{N\to\infty}\vec{F}_{res}^N=\lim_{N\to\infty}\sigma_N\vec S_\alpha=K^{1/\alpha}\vec S_\alpha\;(0<\alpha<1),
\end{equation}

\noindent with $\sigma_N$ being the dispersion and a constant parameter.

\subsection{Exponentially Damped Power Law Force}
\label{appd}

We shall now  consider an inter-particle potential which is short-range, viz.\ decay faster than $r^{-d}$ at long distances,
but its proportional to $r^{-(\delta-1)}$  for short distances. As an example of this class of potentials, let us consider
an exponetially damped power law potential of the form:
\begin{equation}
\label{equ38}
V_i^N(\vec R)= \frac{\kappa_N e^{\displaystyle -\lambda_N |\vec R|} }{|\vec{R}|^{\delta-1}},
\end{equation}   
where $\kappa_N\in\R$, $\lambda_N\ge 0$ and $\delta\ge 0$. 
The force derived from this potential is given by:
\begin{equation}
\label{equ39}
{\vec F}^N_i(\vec R)=\frac{\kappa_N e^{\displaystyle \lambda_N |\vec R|}}{|\vec R|^{\delta}}\left(\lambda_N|\vec R|+\delta-1\right){\hat R}.
\end{equation}
Let observe that when $\lambda_N=0$ we have back a random force like the analysed in the previous section.

Let us define the renormalized random vector
\[
\vec U_{\bar\lambda}=\frac{e^{\displaystyle -\bar\lambda |\vec R|} }{|\vec{R}|^{\delta}}\left(\bar\lambda|\vec R|+\delta-1\right)\hat{R},  
\]
associated to a given real positive number $\bar\lambda$.
Then, it is straightfoward to show that in the assymptotic limit of $|\vec R|\rightarrow 0$ ($|\vec U_{\bar\lambda}|\rightarrow\infty$) the density distribution $\rho_{\vec{U}_{\bar\lambda}}(\vec{u})$ can be written as
\begin{equation}
\label{equ40}
\rho _{\vec U_{\bar\lambda}} \left(\vec{u}\right)=\rho _{\vec{R}} \left(\vec{r}\right)\left\| \displaystyle \frac{\partial \vec{r}}{\partial \vec{u}} \right\|,
\;\; 
\vec u(\vec r)=(\delta-1)\frac{\hat r}{|\vec r|^{\delta}},
\end{equation}
where $\left\|{\partial \vec{u}}/{\partial \vec{r}} \right\|$ is the jacobian of $\vec{u}(\vec{r})$. Analogously as in (\ref{equ24}) we can conclude that
\begin{equation}
\label{equ41}
\lim_{|\vec u|\to \infty }\rho_{\vec{U}_{\bar\lambda}}(\vec{u})=\frac{\delta-1}{\delta|\vec u|^{d+d/\delta}}\lim_{|\vec r|\to 0}\rho_{\vec{R}}(\vec{r}).
\end{equation}
Also we consider the density distribuition $\rho_{\vec R}(\vec r)$ in a neighborhood of $\vec r=0$ as given in equation (\ref{equ25}). Thus, we can conclude that assymptotically for $|\vec u|\rightarrow\infty$ we have (like in equation \ref{equ26}):
\begin{equation}
\label{equ42}
\rho_{\vec{U}_{\bar\lambda}}(\vec{u})=
\frac{\delta-1}{\delta}g(\hat{u})\frac{1}{|\vec u|^{d+(d+\nu)/\delta}},
\end{equation}

Thus, in a equivalent way as shown in the previous section, the density distribution $\rho_{\vec U_{\bar\lambda}}(\vec{u})$ can be put in the following form (analogous to equation \ref{equ27}): 
\begin{equation}
\label{equ43}
\rho_{\vec U_{\bar\lambda}} \left(\vec{u}\right)=\left\{
\begin{array}{l}
{f\left(\vec{u}\right){\rm \; \; \; \; \;}|\vec u|\le u_{c} } 
\\ \\ 
\displaystyle {\frac{C\left(\hat{u}\right)}{|\vec u|^{d+\alpha}} {\rm \; \; \; \; \; }|\vec u|>u_{c} } 
\end{array}\right.\;\; 
\alpha =\frac{d+\nu}{\delta} ;{\rm \; \; \; }C\left(\hat{u}\right)=\frac{ \delta-1}{\delta}g\left(\hat{u}\right)  .
\end{equation} 

The resultant random force is given in (\ref{equ18})
and the force $\vec{F}^N_i$ is a renormalization of $\vec U_{\bar\lambda}$ in analogous way as defined in equation (\ref{equ30}), that is,
\begin{equation}
\label{equ44}
\vec{F}^N_i= \frac{\kappa_N e^{\displaystyle -\lambda_N | L_N\vec R|} }{\left|L_N\vec{R}\right|^{\delta}}
\left(\lambda_N|L_N\vec R|+\delta-1\right)
\hat{R}=a_N\vec U_{\bar\lambda}
\end{equation}
where we have considered
\begin{equation}
\label{equ45}
a_N=\frac{\kappa_N}{L_N^\delta};\;\;\lambda_NL_N=\bar\lambda.
\end{equation}
As in the previous section,
we have renormalized the force of each particle in terms of the variable $\vec U_{\bar\lambda}$.

The Vlasov limit calculation proceed in a fully similar way as done in the previous section and, accordingly, we obtain the same results for the assymptotic characteristic function of the resultant force associated to the different values to the shape exponent $\alpha$. 

\subsection{Scale Renormalization Process}
\label{appe}

Another possible renormalization that is based only on a time and spatial scale transformation in the units in which the force is measured. We call this kind of renormalization a {\it Scale Renormalization}. 

In order to define a Scale Renormalization let remember that the test particle satisfies the second law of Newton:

\begin{equation}
\label{equ46}
m\frac{d^2\vec r}{dt^2}=\sum_{i=1}^N \vec F_i,
\end{equation} 
where $\vec r$ stands for the test particle position and the force $F_i$ is defined in equation (\ref{equ22}) or (\ref{equ39}).

Now, let consider a scale tranformation in time and space given by:
\begin{equation}
\label{equ47}
t=T_Nt_N;\;\vec r=L_N\vec r_N;\; \vec R=L_N\vec R_N.
\end{equation}
We remark that $t_N$ is the reescaled time, $\vec r_N$ is the reescaled position vector of the test particle and $\vec R_N$ is the reescaled random vector associated to the relative position of the {\it i'th} particle with respect to the test particle. Their values may depend on the number of particles $N$ through the parameters $T_N$ and $L_N$, which are called scale parameters. 

Plugging the reescaled variables in (\ref{equ47}) into the second Newton law (\ref{equ46}) we get
\begin{equation}
\label{equ48}
m\frac{d^2\vec r_N}{dt_N^2}=\vec F^N_{res}=\sum_{i=1}^N\vec F_i^N,
\end{equation}
where $\vec F_{res}^N$ denotes the resultant force and the reescaled force $\vec F^N_i$ due to the {\it i'th} particle can be expressed  as
\begin{equation}
\label{equ49}
\vec F_i^N=\kappa\left( \frac{T_N^2}{L_N} \right) 
\vec F_i=a_N\vec U;\;\;
a_N= \kappa\frac{T_N^2}{L_N}, 
\end{equation}
where we have considered $\kappa_N=\kappa$ as a force constant in some system of units. 

Similarly to what was obtained in equation (\ref{equ30}), the force $F^N_i$ is a renormalization of the vector $\vec U$, therefore the limit calculation follows the same steps made for the Vlasov and Fluctuation Limit and yields the same conditions for the renormalization parameter  $a_N$ given in Table \ref{tab2}, but taking into account its new value in Eq. (\ref{equ49}), we obtain new renormalization conditions that are summarized in Table \ref{tab3}.

\begin{table}[h!tb]

\caption{\it The Scale Renormalization Conditions. We have  $\alpha=(d+\nu)/\delta$ and $a_N$ given in Eq.~(\ref{equ49}). Without lost of generality we can choose a system of units where $|\kappa|=1$ and $K$ as an arbitrary positive constant.}
\begin{center}
\begin{tabular}{c}
\hline\hline
\begin{tabular}{ccc}
Fluctuation ($0<\alpha<1$) & Singular ($\alpha=1$) & Vlasov ($1<\alpha<\infty$) \\
($0<d+\nu<\delta$) & ($\delta=d+\nu$) & ($\delta<d+\nu$) \\
\hline\hline\\
$\displaystyle\left(\frac{T_N^2}{L_N} 
\right)^{(d+\nu)/\delta}=\frac{K}{N}$ &
$\displaystyle 
-\frac{T_N^2}{L_N} \ln\frac{T_N^2}{L_N} =\frac{K}{N}$ &
$\displaystyle\frac{T_N^2}{L_N}=\frac{K}{N}$    
\end{tabular}\\ \\\hline\hline
\end{tabular}
\end{center}
\label{tab3}
\end{table}

\subsection{Physical Renormalization}
\label{physrenorm}

We named the whole process as {\it Physical Renormalization} for it provides changes in the physical parameters that defines the system of $N$ particles. It implies that either the force or the system size, or both, must be modified in the renormalization process. Taking into account the renormalization conditions presented in Table \ref{tab3}  and the definition of $a_N$ in Eq.~(\ref{equ30}) we obtain the renormalization relations that are shown in Tables~\ref{tab5} and \ref{tab4}.

\begin{table}[!htb]

\caption{\it The Physical Renormalization Conditions. We have  $\alpha=(d+\nu)/\delta$ and $a_N$ given in Eq.~(\ref{equ30}). $K$ is an arbitrary positive constant. }
\begin{center}
\begin{tabular}{c}\hline\hline
\begin{tabular}{ccc}
Fluctuation ($0<\alpha<1$) & Singular ($\alpha=1$) & Vlasov ($1<\alpha<\infty$) \\
($0<d+\nu<\delta$) & ($\delta=d+\nu$) & ($\delta<d+\nu$) 
\end{tabular}\\\hline\hline \\
\begin{tabular}{ccc} 
$\displaystyle N\frac{|\kappa_N|^{(d+\nu)/\delta}}{L_N^{d+\nu}}=K$ &
$\;\;\;\;\;
\displaystyle -N\frac{|\kappa_N|}{L_N^\delta}
\ln\frac{|\kappa_N|}{L_N^\delta}=K$ &   
$\;\;\;\;\;
\displaystyle N\frac{|\kappa_N|}{L_N^\delta}=K$
\end{tabular}\\ \\\hline\hline
{\it (i) - The Size System is Not Renormalized} ($L_N=1$)\\\hline\hline\\
\begin{tabular}{ccc}
$\displaystyle
|\kappa_N|=\left(\frac{K}{N}\right)^{\delta/(d+\nu)}$ &
$\;\;\displaystyle -|\kappa_N|\ln|\kappa_N|=\frac{K}{N}$ &
$\;\;\;\;\;\;\;\;\;\;\displaystyle |\kappa_N|=\frac{K}{N}$ 
\end{tabular}\\ \\\hline\hline
{\it (ii) - The Force Constant is Not Renormalized} ($|\kappa_N|=1$)
\\\hline\hline\\
\begin{tabular}{ccc}
$\;\;\;\;\displaystyle N=KL_N^{d+\nu}$  & 
$\;\;\;\;\;\;\;\;\;\;\displaystyle N=\frac{K}{d+\nu}\frac{L_N^{d+\nu}}{\ln L_N}$ & 
$\;\;\;\;\;\;\;\;\displaystyle N=KL_N^\delta$ 
\end{tabular}\\ \\\hline\hline
\end{tabular}  
\end{center}
\label{tab4}
\end{table}

There are two particular conditions that are worth to be discussed in detail and are shown in Table~\ref{tab4}: {\it (i)} the size of the system is kept fix and {\it (ii)} the force constant  remains unchanged. Without lost of generality, we may consider either $L_N=1$ or $\kappa_N=1$ to obtain the renormalization conditions.

In the case {\it(i)} the system size is not renormalized and for $\alpha> 1$ the constant force $\kappa_N$ should be the inverse of the number of particles $N$ and this relation is independent of the power law exponent $\delta$ associated to the force. It assures the extensivity of the potencial energy associated to the force, as well known in the Vlasov Limit. In the Fluctuation Limit for $\alpha<1$ the force constant is the inverse of the number of particles trough the power $\alpha=\delta/(d+\nu)$  (as shown in the second line of Table~\ref{tab4} and, differently of the Vlasov Limit, this renormalization does not define an extensive potential energy.  

In the case {\it(ii)}, where the force constant is not renormalized, the Vlasov Limit ($\alpha > 1$) imply a relation between the number of particles $N$ and the size system that depends only on the power law exponent $\delta$ of the force. On the other hand, for the Pure Fluctuation Limit ($\alpha < 1$), this relation depends only on $d+\nu$, that is, the sum of the spatial dimension $d$ and the parameter $\nu$, which is associated to the spatial density distribution of particles in the neighborhood of the test particle (see equation \ref{equ25}).

Particularly, in the Pure Fluctuation Limit, if $\nu=0$ then the relation between $N$ and $L_N$ implies the well known thermodinamical limit: the size and the number of particles goes to infinity while the ratio between the number of particles and the system volume remain constant. Moreover, the random resultant force is only a fluctuation force and the dispersion parameter $\sigma_N=K^{1/\alpha}$ depends on the particle number density.

\section{Discussions}
\label{conclusion}

The results presented in the previous sections are rather mathematical and we need to clarify what are the mathematical constraints imposed in our calculations in order to compare to previous results of the literature (see~\cite{chavanis}). The conditions usually adopted in others results refer to isotropy and homogeneity over the distribution $\rho(\vec r)$. 

Firstly, we have not restricted the exponent $\delta$ of the force acting on the test particle and also no global restriction is applied to distribution of the $N$ particles. Secondly, we just imposed that the limit of the distribution for $r\rightarrow 0$ can be expressed as $\lim_{r\rightarrow 0}\rho(\vec r)=g(\hat v)r^\nu$ with $-d<\nu$, $\delta\geq 0$ e $0<\alpha\leq\infty$.

Along these lines let us discuss in more detail some particular situations. One has: 

\begin{enumerate} 

\item For $\nu=0$  the limit $\lim_{r\rightarrow 0}\rho(\vec r)$ exist and is finite. It encompass all distributions $\rho(\vec r)$ finite at $r=0$ (continuous or not).

\begin{enumerate}[label=(\roman*)]

\item The condition $g(\hat r)=C>0$ represents a distribution $\rho(\vec r)$  that is finite ($\rho(0)=C>0$) and continous at $r=0$. In this case, the assymptotic distribution of the resultant force is given by a stable and symmetric L\'evy distribution for $0<\alpha\leq 2$ and a Gaussian distribution for $\alpha>2$.

\item If $g(\hat r)$ is not constant then the distributions are finite but not continous at $r=0$. In this case the assymptotic value for the resultant force may be represented by stable and assymmetric L\'evy distributions  for $0<\alpha<2$ and for Gaussian distributions for $\alpha\geq 2$. The parameters that define these distributions: of form ($A_{\alpha}(\hat z)$) and assimetry  ($\beta_{\alpha}(\hat z)$) are completely determined by the function $g(\hat r)$ (see Eqs.~(\ref{equacao5}))~and~(\ref{equ29}). This corroborates the result of \cite{kandrup} for the $3d$ gravitational force. It means that the knowldege of the distribution $\rho(\vec r)$ nearby the test particle (at $r=0$) is sufficient to determine the assymptotic expression for the resultant force.

\end{enumerate}

\item The condition $-d<\nu<0$ represents a class of distributions $\rho(\vec r)$ that are singular at $r=0$.  For $\nu>0$ one has continuous distributions ($\rho(\vec r)$) at $r=0$ such as $\rho(0)=0$, but some first or higher order derivative is singular at $r=0$. The distribution of the resultant force (assymptotic) is always Gaussian for $\alpha\geq 2$ and a stable L\'evy for $0<\alpha\leq 2$. For the latter, the distribution can be symmetric for $g(\hat r)=C>0$, os assymmetric when $g(\hat r)$ is not constant.

\end{enumerate}

Now we should explore the limit $N\rightarrow\infty$ for the resultant force. In this situation one has defined two limits: {\it Fluctuation Limit} and {\it Vlasov Limit}. In both cases we must renormalize either constant of the force ($\kappa_N$) or the size of the system ($L_N$) or both.

\subsection{Fluctuation Limit}

The Fluctuation Limit for $0<\alpha<1$ may be defined without any further mathematical constraint on the distribution $\rho(\vec r)$ other than those indicated at the introduction of this section (See $1^{st}$~column,~Tables~\ref{tab2}~and~\ref{tab3}).  For $\alpha\geq 1$ one has to impose a further constraint to the distribution $\rho(\vec r)$, i.e., the average value for the resultant force must be zero for $\alpha> 1$ and $g(\hat r)$ must be symmetric when $\alpha=1$.

 We should also notice that in the case of the Central Limit Theorem (CLT)(see Section \ref{clt}) the resultant force is central (with the same meaning of CLT) if one has a non null average for the resultant force. This procedure is not possible in the Fluctuation Limit for one is allowed only to renormalize either the constant of the force or the size of the system, or both. However, when the average of the resultant force is zero this procedure of centralization is not needed in the CLT and the Fluctuation Limit is equivalent to the renormalization process of the CLT discussed in Sec.~\ref{LK} for the conditions presented in Table~\ref{tab5}. Some additional considerations related to the results presented in the literature are necessary. They are: 

\begin{enumerate}[label=(\roman*)]

 
\item The results obtained by Chavanis~\cite{chavanis} are reproduced in the context of the fluctuation limit when $\kappa_N$ is constant and the size of the system ($L$) increases with $N\rightarrow\infty$ (see case(ii) Table~\ref{tab5}). In~\cite{chavanis} the distribution is isotropic and $0<\alpha<2$ and these are not needed here. For $1\leq \alpha<2$ also referring to~\cite{chavanis} we obtain the same result imposing that the average force is zero but not imposing isotropy. The condition $2\leq\alpha<\infty$ (column $3$ of Table~\ref{tab5}) is not analized by~\cite{chavanis}.
 
 \item The $2d$ system studied by Chavanis~\cite{chavanis} means to consider  $\delta=1$ and $d=2$ for an uniform distribution around the test particle. In this case the thermodynamic limit does not exist but one can obtain a well defined fluctuaction limit if the relations given in column $2$ of Table~\ref{tab5} are satisfyied. It means that the result obtained in~\cite{chavanis} is an assymptotic relation for the resultant force and the number of particles in the system. In the context of this report it is given by the assymptotic equations obtained from the CLT contained in Eq.~\ref{equ15} for $\alpha=2$, Sec.~\ref{LK}, Table \ref{tab1}.
 
 \item In ~\cite{chavanis} Chavanis extends the study to encompass $\delta=0$ and $d=1$. He admits a symmetric distribution of particles around the test particle that means $\alpha=\infty$. From the third column of case (ii) of Table~\ref{tab5} the fluctuation renormalization is not permitted for the number of particles is independent of the size of the system. However, according to the relations given in column 3 case (ii) of Table~\ref{tab5} if the constant of the force depends on the number of particles one may have a fluctuation renormalization. In the context of this report the result presented in \cite{chavanis} (an assymptotic relation for the resultant force obtained from the CLT) may be described by the condition $\alpha>2$ (Gaussian Distribution, see Table~\ref{tab1}).

\item For $\nu=0$ and the force constant not renormalized (see case(ii) in Tables~\ref{tab4} and \ref{tab5}) but renormalizing the size of the system ($L(N)$) one has for $0<\alpha<2$ the well know {\it Thermodynamic Limit}. The results of Chandresakhar~\cite{chandra} may be obtained if one takes $\delta=2$ and $d=1$ corresponding to a limit distribution of resultant force given by a L\'evy with $\alpha=3/2$. In this context the hypothesis of uniformity of~\cite{chandra} for the distribution means that the average of the force is null.

\item The results for the resultant force obtained in Ahmed et.~al.~\cite{ahm} (system with fixed size and $N\rightarrow\infty$) are derived from the CLT. In the context of the present report this is a particular case of  $1<\alpha< 2$ in Eq.~\ref{equ15} taking $\alpha=3/2$ and $<\vec U>=0$.

\end{enumerate}

To conclude the discussion of this limit we should emphasize that in the literature those reports that investigate the resultant random force follow the same methology of the CLT and we have not found any calculation that resembles the limit processes such as the constant of the force is renormalized. We provide in the Table~\ref{tab5} the conditions for renormalization in this limit for $1\leq\alpha<\infty$.

\begin{table}[!htb]

\caption{\it  Fluctuation Limit for $1\leq\alpha=(d+\nu)/\delta<\infty$. The condition $\left<{\vec F}_N\right>=N\left<\vec U\right>=0$ must be satisfied and $K>0$ is an arbitrary constant.}
\begin{center}
\begin{tabular}{c}\hline\hline  Fluctuation Renormalizations \\ \hline\hline
\begin{tabular}{ccc}
\\
$1\leq\alpha<2$ & $\alpha=2$ &  $2<\alpha<\infty$  \\
\hline\hline\\
$N|a_N|^\alpha=K$ & $-N|a_N|^2\ln|a_N|=K$ &  $N|a_N|^2=K$ \\ \\
$\displaystyle N\frac{|\kappa_N|^{(d+\nu)/\delta}}{L_N^{d+\nu}}=K$ &
$
\displaystyle -N\frac{|\kappa_N|^2}{L_N^{2\delta}}
\ln\frac{|\kappa_N|}{L_N^\delta}=K$ &   
$
\displaystyle N\frac{|\kappa_N|^2}{L_N^{2\delta}}=K$
\end{tabular}\\ \\\hline\hline
{\it (i) - The Size System is Not Renormalized} ($L_N=1$)\\\hline\hline\\
\begin{tabular}{ccc}
$
\hspace{1mm}\displaystyle
|\kappa_N|=\left(\frac{K}{N}\right)^{\delta/(d+\nu)}$ &
$\displaystyle -|\kappa_N|^2\ln|\kappa_N|=\frac{K}{N}$ &
$\displaystyle |\kappa_N|=\left({\frac{K}{N}}\right)^{1/2}$ 
\end{tabular}\\ \\\hline\hline
{\it (ii) - The Force Constant is Not Renormalized} ($|\kappa_N|=1$)
\\\hline\hline\\
\begin{tabular}{ccc}
$\;\;\;\;\displaystyle N=KL_N^{d+\nu}$  & 
$\;\;\;\;\;\;\;\;\;\;\displaystyle N=K\frac{L_N^{2\delta}}{\delta\ln L_N}$ & 
$\;\;\;\;\;\;\;\;\displaystyle N=KL_N^\delta$ 
\end{tabular}\\ \\\hline\hline
\end{tabular}  
\end{center}
\label{tab5}
\end{table}

\subsection{Vlasov Limit}

The Vlasov limit is well defined if the average of the resultant force exists and it means that one must consider $1\leq\alpha$. In this situation no restrictive hypothesis is necessary for the distribuition $\rho(\vec r)$ of particles around the test particle. The limit distribution of the force for $N\rightarrow\infty$ will be given by a Dirac $\delta$-function and the assymptotics expressions show that the average of the distribution of the force for $N>>1$ are stables functions of L\'evy (either symmetric or assymmetric) when $1\leq\alpha<2$ or Gaussian distributions for $2\leq\alpha<\infty$ (see Eqs.~\ref{equ33}~and~\ref{equ34}). 

On one side, in the literature the Vlasov limit is obtained for system with constant size. This is shown in columns $2$ and $3$, case (i) of Table~\ref{tab4} of the present report and corresponds to what is known as Kac prescription: the renormalization of the constant of the force (or of the potential energy)
is given by $|\kappa_N|=K/N$. On the other side according to colums $2$ and $3$ of case (ii) of Table~\ref{tab4} one has an opposite situation, or that the constant of the force is fixed and the size of the system varies as a function of $N$. Up to our knowledge this is not discussed in the literature. According to Eqs.~\ref{equ33}~and~\ref{equ34} the distribution of the resultant force, whatever renormalization is effected, one finds the same dependency with $N$. In fact, once we have a value of $\alpha$ one may admit that test particles under the effect of a same average resultant force and of a same fluctuation around this mean value of the force  will have the same dynamics independent of the process of renormalization.
Last, but not least we know that the Vlasov equation is a kinetic equation in the limit of $N\rightarrow\infty$ and obtained in the study Hamiltonian systems within the scope of statistical mechanics. In the literature, it is usual for a gravitational force, that the Vlasov Limit is obtained for a system of $N$ identical particles of mass $m$ with a given spatial distribution and one takes the limit $N\rightarrow\infty$ with $M=Nm$ being constant.

\section{Acknowledgments}

ADF and TMRF acknowlodge CNPq., Brazilian Government for the financial support. The authors would like to acknowledge the frutiful comments of the referee.

\section*{Appendices}

In order to make the paper self contained we include a set of appendices related to theory developed in the main text.

\appendix
\numberwithin{equation}{section}
\section{The Limit Characteristic Function of the Renormalization Process}
\label{appb}

For $0<\alpha<1$ the following renormalization ensures the convergence of the renormalized sequence:   
\begin{equation}
a_N=\frac{1}{N^{1/\alpha}},\hspace{5mm}\vec{b}_N=0.
\end{equation}
From Eqs.~(\ref{equ11}--\ref{equ13}) we obtain:
\begin{eqnarray}
\label{equacao13}
\lefteqn{\psi _{\bar{F}^{N}_{res} } \left(\vec{z}\right)=\psi_{\vec{U}}\left(\frac{\vec{z}}{N^{1/\alpha}}\right)^{N}=e^{-|\vec{z}|^\alpha}}
\nonumber\\
 & & \times\exp\left(
\frac{\pi}{\Gamma(\alpha+1)}\frac{\cos(\alpha\pi/2)A_\alpha(\hat{z})-i\sin(\alpha\pi/2)B_\alpha(\hat{z})}{\sin(\alpha\pi)}+\Omega\left(\frac{\vec{z}}{N^{1/\alpha}}\right)\right).
\nonumber\\
 & & 
\end{eqnarray}
In the $N\to \infty$ limit we have that:
\begin{equation}
\label{equacao14}
\lim_{N\to\infty}\Omega\left(\frac{\vec{z}}{N^{1/\alpha}}\right)=\Omega\left(0\right)=0.
\end{equation}
Thence we have for $\Phi(\vec{z})$ in Eq.~(\ref{equ14}):
\begin{equation}
\label{equacao15}
\Phi\left(\vec{z}\right)=\exp\left(
-\left|\vec{z}\right|^{\alpha }\frac{\pi}{\Gamma(\alpha+1)}\:\frac{\cos(\alpha\pi/2)A_\alpha(\hat{z})-i\sin(\alpha\pi/2)B_\alpha(\hat{z})}{\sin(\alpha\pi)}\right).
\end{equation} 

For the interval $1<\alpha<2$ we have that $\langle\vec{F}^N_{res}\rangle=N\langle\vec{U}\rangle$ and we chose a renormalization such that the renormalized
variable $\bar{F}^N_{res}$ has zero mean and dispersion parameter not increasing with $N$:
\begin{equation}
a_N=\frac{1}{N^{1/\alpha}},\hspace{5mm}\vec{b}_N=-\frac{\langle\vec{F}^N_{res}\rangle}{N^{1/\alpha}},
\end{equation}
The characateristic function os $\bar{F}^N_{res}$ is then:
\begin{equation}
\Phi_{\bar{F}^N_{res}}(\vec{z})=\exp\left(-i\frac{N\vec{z}\cdot\langle\vec{U}\rangle }{N^{1/\alpha}}\right)\:\left[\psi _{\vec{U}}\left(\frac{\vec{z}}{N^{1/\alpha}}\right) \right]^{N}.
\label{alpha12}
\end{equation}
Using the expression for $\psi_{\vec{U}}\left(\vec{z}\right)$ from Eqs.~(\ref{equ4}), (\ref{equ5}) and~(\ref{equ6}) in Eq.~(\ref{alpha12})
we obtain the same expression as in Eq.~(\ref{equacao13}). Taking the limit $N\to\infty$ and considering Eq.~(\ref{equacao14}) we obtain the same result as in Eq.~(\ref{equacao15}). 

For $\alpha=1$ the convenient renormalizations is:
\begin{equation}
a_N=\frac{1}{N},\hspace{5mm}\vec{b}_N=-\langle\vec U\rangle_{u_c}+\vec v\left(1-\gamma-\ln u_c\right)-\vec v\ln N.
\end{equation}
The characteristic function in Eq.~(\ref{equacao13}) and the expression for $\psi_{\vec{U}}\left(\vec{z}\right)$ in Eqs.~(\ref{equ4}) and~(\ref{equ7}) imply that 
\begin{eqnarray}
\lefteqn{\psi _{\bar{F}_{res}^N}(\vec{z})=\psi_{\vec{U}} \displaystyle\left(\frac{\vec{z}}{N}\right)^N}
\nonumber\\
& & =\exp\left(
-|\vec{z}|\left\{\frac{\pi}{2}A_1(\hat z)-i[\hat z\cdot\langle U\rangle_{u_c}+B_{1}(\hat z)(1-\gamma -\ln u_c)
\right.\right.
\nonumber\\
 & & \left.\left.
\hspace{15mm}+B_1(\hat z)\ln\frac{|\vec z|}{N}-D_1(\hat z)]
+\Omega\left(\frac{\vec{z}}{N}\right)\right\}\right)
\end{eqnarray}
Taking $N\to\infty$ in the last equation yields:
\begin{equation}
\label{equacao16}
\psi _{\bar{F}_{res}^N}(\vec{z})=
\exp\left(
-|\vec{z}|\left\{\frac{\pi}{2}A_1(\hat z)
-\hat z\cdot\vec v\ln|\vec z|-D_1(\hat z)\right\}\right)
\end{equation}
where
\begin{equation}
\label{equacao16a}
\vec{v}=-\int _{S_d}C(\hat u)\hat{u}\:dS_{d}. 
\end{equation} 
Here we have considered the following simplification obtained from the definition of $B_1\left(\hat{z}\right)$ in Eq.~(\ref{equacao5}):
\begin{equation}
\label{equacao17}
B_{1} \left(\hat{z}\right)=\int _{S_{d} }C\left(\hat{u}\right)\hat{z}\cdot \hat{u}dS_{d} =
\hat{z}\cdot\left(
\int _{S_{d}}C\left(\hat{u}\right)\hat{u}dS_{d}\right)= 
-\hat{z}\cdot \vec{v}\;.
\end{equation} 
Let remember that the definition of $D_1(\hat z)$ is also given in Eq.~(\ref{equacao5}). 

The characteristic function given in Eq.~(\ref{equacao16}) does not represent a true stable probability distribution, provided that the function $C(\hat u)$ defined in (\ref{equ2}) is not symmetric in the hypersphere $S_d$. In this case we have that $\vec v\neq 0$ and $D_1(\hat z)$ and the respective non symmetric probability distribution is called semi-stable. The semi-stable probability densities of real random variable was first introduced by Levy \cite{levy2}. On the other side, if the function $C(\hat u)$ is symmetric, that is, $C(\hat u)=C(-\hat u)$, then $\vec v=0$ and $D_1(\hat z)=$ and the characteristic function in (\ref{equacao16}) corresponds to a symmetric stable probability distribution which constitutes a generalization of the Cauchy 
distribution of a real random variable to a $d$-dimensional random vector.

For $\alpha=2$ the proper renormalization is:
\begin{equation}
a_N=\frac{1}{\left(N\ln N\right)^{1/2}}\hspace{5mm}
\vec{b}_N=-\frac{\langle\vec{S}_N\rangle}{\left(N\ln N\right)^{1/2}},
\end{equation}
where $\langle\vec{F}_{res}^N\rangle=N\langle\vec{U}\rangle$. Thus, the characteristic function in (\ref{equacao13}) can be written as 
\begin{equation}
\psi_{\bar{F}_{res}^N}(\vec{z})=\exp\left(-i\frac{N\vec{z}\cdot\langle\vec{U}\rangle }{\sqrt{N\ln N}}\right)\psi_{\vec{U}}\left(\frac{\vec{z}}{\sqrt{N\ln N}}\right)^N, 
\end{equation}
and thence:
\begin{eqnarray}
\lefteqn{\psi _{\bar{F}_{res}^N}(\vec{z})=\exp\left(-\frac{\left|\vec{z}\right|^{2}}{2\ln N}\right.}
\nonumber\\
 & &
\times\left[
\hat z\cdot\bar{M}_{u_c}\cdot\hat z+A_2(\hat z)\left(\frac{3}{2}-\gamma-\ln u_c-\ln\left|\frac{\vec{z}}{\sqrt{N\ln N}}\right|\right)\right.
\nonumber\\
 & & \left.\left.
\hspace{10mm}-D_2(\hat z)+i\frac{\pi}{2}B_2(\hat z)+\Omega\left(\frac{\vec{z}}{\sqrt{N\ln N}}
\right)\right]\right),
\end{eqnarray}
where we used the corresponding characteristic function $\psi_{\vec{X}}(\vec{z})$ for $\alpha=2$ givev in Eqs.~(\ref{equ4}) and~(\ref{equ8}).
Finally, in the $N\to\infty$ limit we obtain:
\begin{equation}
\label{equacao19}
\Phi\left(\vec{z}\right)=e^{\displaystyle
-\frac{1}{4} A_{2} \left(\hat{z}\right)\left|\vec{z}\right|^{2} } =e^{\displaystyle
-\frac{1}{2} \left|\vec{z}\right|^{2} \hat{z}\cdot{\bf M}\cdot\hat{z}}
=e^{\displaystyle
-\frac{1}{2}\vec{z}\cdot{\bf M}\cdot\vec{z}},
\end{equation} 
where we used:
\begin{equation}
\label{equacao20}
A_2 (\hat{z})=\int _{S_{d}}C(\hat{u})\:(\hat{z}\cdot\hat{u})^2\: dS_{d}
=\int _{S_d}C\left(\hat{u}\right)\left(\sum_{k=1}^{d}[\hat{z}]_k[\hat{u}]_k\right)^2 dS_{d}=2\hat{z}\cdot{\bf M}\cdot\hat{z},    
\end{equation}
with
\begin{equation}
\label{equacao20a}
\left[M\right]_{ij} =\frac{1}{2} \int _{S_{d} }\left[\hat{u}\right]_{i} \left[\hat{u}\right]_{j} C\left(\hat{u}\right)dS_{d},
\end{equation}
and
\begin{equation}
\hat{u}=([\hat{u}]_1,\ldots,[\hat{u}]_d),\hspace{5mm}\hat{z}=([\hat{z}]_1,\ldots,[\hat{z}]_d).
\end{equation}

Let us now turn to the remaining case $2<\alpha<\infty$. In this case we take:
\begin{equation}
a_N=\frac{1}{N^{1/2}},\hspace{5mm}\vec{b}_N=-\frac{\langle\vec{S}_N\rangle}{N^{1/2}},
\end{equation}
where again $\langle\vec{F}_{res}^N\rangle=N\langle U\rangle$. Plugging the values of $a_N$ and $\vec{b}_N$ into Eq.~(\ref{equacao13}) yields:
\begin{equation}
\psi_{\bar{F}_{res}^N}(\vec{z})=\exp\left(-i\frac{N\vec{z}\cdot\langle\vec{U}\rangle}{\sqrt{N}}\right)\psi_{\vec{U}}\left(\frac{\vec{z}}{\sqrt{N}}\right)^N. 
\end{equation}
Using Eqs.~(\ref{equ4}) and~(\ref{equ9}) we obtain:
\begin{equation}
\psi_{\bar{F}_{res}^N}(\vec{z})=\exp\left(
-\frac{1}{2}|\vec{z}|^2\left[\hat{z}\cdot\bar{M}\cdot\hat{z}+\Omega\left(\frac{\vec{z}}{\sqrt{N}}\right)\right]\right).
\end{equation}
Finally, taking the limit $N\rightarrow\infty$ leads to:
\begin{equation}
\label{equacao19a}
\Phi\left(\vec{z}\right)=\exp\left(-\frac{1}{2} |\vec{z}|^{2}\hat{z}\cdot\bar{M}\cdot\hat{z}\right)=\exp\left(-\frac{1}{2}\vec{z}\cdot\bar{M}\cdot\vec{z}\right), 
\end{equation}
where $\bar{M}$ is the covariance matrix in Eq.~(\ref{equ1}).

\section{Some Properties of L\'evy-Khintchine Distribution}
\label{appc}

It is worth to emphasize some points about the L\'evy-Khintchine characteristic function of a $n-dimensional$ random vector as defined in (\ref{equ15}):
\begin{enumerate}
\item This Characteristic Function is specified by a real parameter $\alpha\in (0,2]$, called in this work shape-exponent, and by a non-negative function $C\left(\hat{x}\right)$, defined in the hypersphere of unit radius contained in $\R^n$, that is, $\hat{x}\in S_n\subset\R^n$. Additional conditions on this function must be done only in order to garantee the existence of the integrals defined in (\ref{equacao5}). 
\item The stability of the L\'evy-Khintchine distribution is related to the following property: Let consider two random vectors $\vec{X}_1$ and $\vec{X}_2$ having characteristic function as given in (\ref{equ15}) with the same $\alpha$ parameter and $C(\hat{x})$ function, and having different parameters $\lambda$: let say $\lambda_1$ and $\lambda_2$ respectively associated to $\vec{X}_1$ and $\vec{X}_2$. Then, the characteristic function associated to the random vector $a_1\vec{X}_1+a_2\vec{X}_2$ (a linear combination of $\vec{X}_1$ and $\vec{X}_2$ with $a_1$ and $a_2$ being real numbers) will have the same form, but with $\lambda$ parameter given by 
$\lambda_1a_1^{\alpha}+\lambda_2a_2^{\alpha}$.
\item From the function $C\left(\hat{x}\right)$ we define the
functions $A_\alpha\left(\hat{z}\right)$ and $B_\alpha\left(\hat{z}\right)$ through the formulas in (\ref{equ15}). Let observe that these functions depends only on the direction $\hat{z}$ associated to a given vector $\vec{z}\in\R^n$. It is straightfoward to show that $A_\alpha\left(\hat{z}\right)$ is a simetric and $B_\alpha\left(\hat{z}\right)$ an anti-simetric function. Mathematicaly we have: 
$A_\alpha\left(-\hat{z}\right)=A_\alpha\left(\hat{z}\right)$ and
$B_\alpha\left(-\hat{z}\right)=-B_\alpha\left(\hat{z}\right)$.
\item The function $A_\alpha\left(\hat{z}\right)$ is positive for any $\hat{z}$ and geometrically it can be associated to a width measure of the density probability function in a certain direction. 

The function $\beta_\alpha\left(\hat{z}\right)$ is called asymmetry function and, from the inequality 
$\left|B_\alpha\left(\hat{z}\right)\right|
\le A_\alpha\left(\hat{z}\right)$ for all $\hat{z}$ we have $-1\le \beta_\alpha\left(\hat{z}\right)\le 1$. From the property of asymmetry of the function $B_\alpha(\hat{z})$ it is easy to conclude that the asymmetry function is also an asymmetric function, that is, $\beta_\alpha(-\hat{z})=-\beta_\alpha(\hat{z})$.
\item
In order to have a better understanding about the meaning of the functions $A_\alpha(\hat{z})$ and $\beta(\hat{z})$ let consider the marginal distribuition 
\[\displaystyle
\rho_k(x_k)=\int\cdots\int\rho\left(x_1,\ldots,x_k,\ldots,x_n\right)
dx_1\ldots dx_{k-1}dx_{k+1}\ldots dx_n
\] 
of the $k'th$ random component $X_k$ of the random vector 
\[
\vec{X}=\left(X_1,\ldots,X_k,\ldots,X_n\right),
\]
which is supposed to have a L'evy-Khintchine characteristic function as given in (\ref{equ15}). 

It is straightfoward to show that the
characateristic function $\psi_k(z_k)$ associated to the marginal distribution $\rho_k(x_k)$ can be obtained just taking $\vec{z}=z_k\hat{e}_k$   in equation (\ref{equ15}), where $\hat{e}_k$ stand for the canonical versor with components $\left(\hat{e}_k\right)_i=\delta_{ki}$ for $i=1,\ldots,n$. In this case we need to consider that $\hat{z}=\hat{e}_k$ to obtain:
\[
\psi_k\left(z_k\right)=\Psi\left(z_k\hat{e}_k\right)=
e^{\displaystyle
-\frac{\lambda A_k\left|z_k\right|^{\alpha}}{\Gamma\left(\alpha+1\right)}\left(1-i\frac{z_k}{\left|z_k\right|}\beta_k
\tan\left(\frac{\alpha\pi}{2}\right)\right)};
\]
where we have defined 
$A_k=A_{\alpha}\left(\hat{e}_k\right)$
$\beta_k=\beta_{\alpha}\left(\hat{e}_k\right)$.
The characteristic function above has the well known form of the L\'evy-Khintchine characteristic function associated to one-dimensional stable random variables \cite{levykhintchine}.

This shows that the marginal density probability of any component of a stable random vector corresponds to a stable random variable
with shape exponent $\alpha$ and asymmetry parameter $\beta_k$. The parameter given by $\lambda A_k$ can be used as a measure of the width of the marginal distribution $\rho_k(x_k)$.            
\end{enumerate}

\end{document}